\documentclass[sigconf]{acmart}
\usepackage{multirow}
\usepackage{xcolor}
\usepackage{soul}
\usepackage[utf8]{inputenc}
\usepackage{subcaption}
\usepackage{xfrac}
\usepackage{fancybox}
\usepackage{float}

\AtBeginDocument{%
  }

\copyrightyear{2026}
\acmYear{2026}
\setcopyright{cc}
\setcctype{by-nc-nd}
\acmConference[CHI '26]{Proceedings of the 2026 CHI Conference on Human Factors in Computing Systems}{April 13--17, 2026}{Barcelona, Spain}
\acmBooktitle{Proceedings of the 2026 CHI Conference on Human Factors in Computing Systems (CHI '26), April 13--17, 2026, Barcelona, Spain}
\acmPrice{}
\acmDOI{10.1145/3772318.3790522}
\acmISBN{979-8-4007-2278-3/2026/04}

\begin{document}
\setlength{\tabcolsep}{0.8pt}

\title[NSFW (Not-Safe-For-Work) Chatbots on FlowGPT]{When Generative AI Is Intimate, Sexy, and Violent: Examining Not-Safe-For-Work (NSFW) Chatbots on FlowGPT}

\author{Xian Li}
\authornote{Both authors contributed equally to this work and share first authorship.}
\email{xli359@connect.hkust-gz.edu.cn}
\orcid{0009-0005-7563-503X}
\affiliation{%
  \institution{Hong Kong University of Science and Technology (Guangzhou)}
  \streetaddress{}
  \city{Guangzhou}
  \state{}
  \country{China}
  \postcode{}
}

\author{Yuanning Han}
\authornotemark[1]
\email{hany875@newschool.edu}
\orcid{0009-0002-4978-8532}
\affiliation{%
  \institution{Parsons School of Design}
  \streetaddress{}
  \city{New York}
  \state{New York}
  \country{USA}
  \postcode{}
}

\author{Di Liu}
\email{seucliudi@gmail.com}
\orcid{0009-0001-3513-6587}
\affiliation{%
  \institution{School of Design, SUSTech}
  \streetaddress{}
  \city{Shenzhen}
  \state{}
  \country{China}
  \postcode{}
}

\author{Pengcheng An}
\email{anpc@sustech.edu.cn}
\orcid{0000-0002-7705-2031}
\affiliation{%
  \institution{School of Design, SUSTech}
  \streetaddress{}
  \city{Shenzhen}
  \state{}
  \country{China}
  \postcode{}
}
\author{Shuo Niu}
\email{shniu@clarku.edu}
\orcid{0000-0002-8316-4785}
\affiliation{%
  \institution{Clark University}
  \streetaddress{950 Main Street}
  \city{Worcester}
  \state{Massachusetts}
  \country{USA}
  \postcode{01610}
}

\renewcommand{\shortauthors}{Li et al.}

\begin{abstract}
\begin{center}
\textbf{{\color{red} Content Warning: This paper contains sexually explicit and violent images and text.}}
\end{center}
User-created chatbots powered by generative AI offer new ways to share and interact with Not-Safe-For-Work (NSFW) content. However, little is known about the characteristics of these GenAI-based chatbots and their user interactions. Drawing on the functional theory of NSFW on social media, this study analyzes 376 NSFW chatbots and 307 public conversation sessions on FlowGPT. Findings identify four chatbot types: roleplay characters, story generators, image generators, and do-anything-now bots. AI Characters portraying fantasy personas and enabling hangout-style interactions are most common, often using explicit avatar images to invite engagement. Sexual, violent, and insulting content appears in both user prompts and chatbot outputs, with some chatbots generating explicit material even when users do not create erotic prompts. In sum, the NSFW experience on FlowGPT can be understood as a combination of virtual intimacy, sexual delusion, violent thought expression, and unsafe content acquisition. We conclude with implications for chatbot design, creator support, user safety, and content moderation.

\end{abstract}



\maketitle

\section{Introduction}
NSFW (Not Safe For Work) is an internet tag used to label online content considered inappropriate for public or professional settings, typically due to explicit sexual or violent material shared on social media. 68.4\% of adolescents in the U.S. have encountered such content online \cite{Jiang2019}. NSFW content remains controversial due to its potential harm to vulnerable users. However, sharing NSFW content also fosters a community of users who use the term to contour their unique interests and identities \cite{paasonen_nsfw_nodate}. In response, research has examined NSFW content and its creators on platforms such as Tumblr \cite{tiidenberg_not_2019, tiidenberg_profiling_2012, Tiidenberg_how_2013}, Instagram \cite{pilipets_247_2024}, and Reddit \cite{matias_going_2016}. More recently, NSFW creators have begun integrating generative AI (GenAI) and large language model (LLM) technologies to deliver new forms of NSFW content \cite{wei2024exploring, wang2024exposure}. One emerging online community fostering this development is FlowGPT.

\begin{figure*}[t!] 
\centering
\includegraphics[width=1\textwidth]{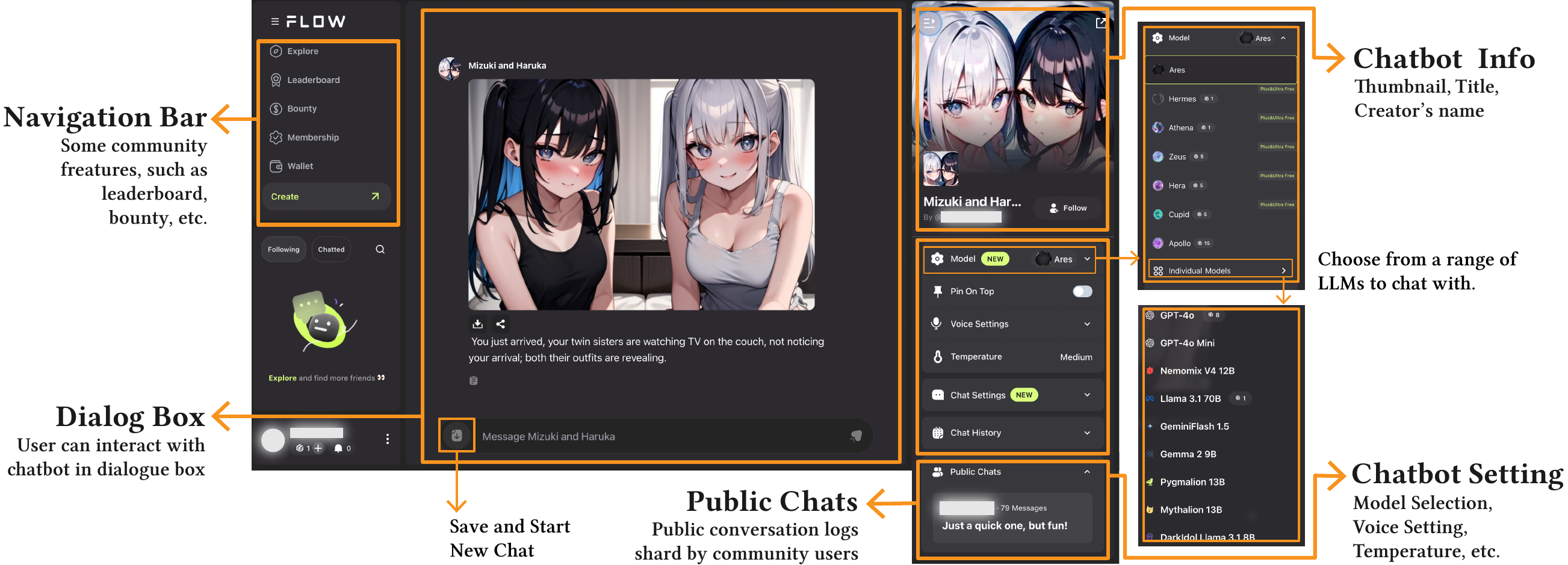} 
\caption{FlowGPT Interface}
\label{fig:flowgpt_interface}
\end{figure*}

\par

\par
Established on February 7, 2023\footnote{\url{https://www.crunchbase.com/organization/flowgpt}}, FlowGPT is an online platform for sharing user-customized large language model (LLM) chatbots. The site presents itself as \textit{``a community for anyone to share and discover the best prompts to unleash the boundless possibilities of Artificial Intelligence''},\footnote{\url{https://flowgpt.com/about}} positioning itself as a central hub for discovering and distributing generative AI chatbots. Since its launch, the platform has attracted millions of daily visitors. As of July 2024, it recorded 4.18 million visits\footnote{\url{https://www.semrush.com/website/flowgpt.com/overview/}}. FlowGPT's design emphasizes a community-centric approach, allowing users to share and explore a wide range of GenAI chatbots and use cases. These chatbots, primarily created by individual users, leverage GenAIs to provide interactive experiences in a dialog box format (see \autoref{fig:flowgpt_interface}). The platform features GenAI-embodied characters from cultural or fictional niches, echoing research showing that fan communities often subvert or reinterpret source material \cite{evans2017more, barnes_fanfiction_2015}. Compared to other AI chatbot platforms such as the GPT Store, FlowGPT offers a more open ecosystem -- granting creators greater flexibility in customizing different GenAI models and enabling users to publicly share their conversations with chatbots.
\par
However, while FlowGPT is not designed for sharing NSFW content, chatbots labeled as ``NSFW'' hold a significant presence on the platform. These chatbots differ from traditional NSFW content on other social media in several key ways. FlowGPT lowers the barrier for creating NSFW AI content by allowing users to utilize GenAI models and design customized prompts, enabling the generation of more natural and interactive language. Accessing NSFW content on FlowGPT requires prompting chatbots, in contrast to consuming NSFW content created by human users on other platforms. These chatbots can even simulate companionship or romantic relationships with users \cite{depounti_ideal_2023}. Despite moderation efforts by GenAI and LLM service providers \cite{ghafouri2023ai}, creators can still modify chatbots and bypass restrictions using jailbroken prompts \cite{liu_jailbreaking_2024, shen_anything_2024, li_flowgpt_2024}, potentially enabling covert production and distribution of explicit content. Moderating this user-generated NSFW chatbots -- complicated by community resistance, technological workarounds, and governance challenges \cite{pilipets_247_2024, Greyson2022Moderation} -- requires a deeper understanding of this emerging online phenomenon.
\par



\par

Previous research on NSFW content in online communities has primarily focused on how the explicit texts and images are distributed and moderated \cite{Krishnappa2024NSFW, tiidenberg_sex_2021, Jeanna2022NSFW}, as well as the motivations behind their production \cite{Tiidenberg_how_2013, hamilton_nudes_2023}. A significant gap remains in understanding GenAI-powered NSFW chatbots. Technosexuality research in HCI calls for critical engagement with how emerging technologies mediate desire and pleasure \cite{Kannabiran2011Sexuality}. Using FlowGPT as a case study, this work provides an understanding of this emerging space and outlines design and moderation implications for platform policymakers, creators, and users engaged in AI-mediated sexuality. It also prepares HCI and CSCW researchers for the growing use of GenAI-powered chatbots as a new medium for delivering NSFW content.
We base our analysis on Paasonen et al.'s conceptualization of NSFW~\cite{paasonen_nsfw_nodate} -- ``NSFW'' as \textit{boundary work}, \textit{an engagement device}, and \textit{a framing device}. As \textit{boundary work}, Paasonen clarifies the tag's various uses and functions to show how it draws boundaries of acceptability. As an \textit{engagement device}, NSFW not only warns of sensitive material but also simultaneously invites certain encounters with explicit content. As a \textit{framing device}, Paasonen argues that NSFW enables analysis of NSFW content dynamics, highlighting the changing contours of risk, safety, and work shaped by digital media.
\par
However, this framework primarily focuses on NSFW content created by human users and within online user communities. It requires re-examination in the context of FlowGPT, where NSFW content circulates through material dynamically generated by GenAI, and where users engage in conversational interactions with chatbots that present both intelligence and explicitness. To address this gap, we analyzed 376 NSFW chatbots and 307 public conversation sessions on FlowGPT.
We categorize chatbot types to examine how NSFW operates as \textit{boundary work} on FlowGPT (i.e., how its uses and functions draw boundaries of acceptability). To analyze NSFW as \textit{an engagement device} (how NSFW serves as an invitation to NSFW encounters), we investigate the identity settings, behavioral traits in initial conversations, and avatar settings used by NSFW chatbots to engage users. To explore NSFW as \textit{a framing device} (how NSFW shapes the dynamics of risks and safety), we identify potentially harmful content present in user-shared conversations with these chatbots. Specifically, we address three research questions:

\begin{itemize}
    \item RQ1: What are the functions of chatbots marked with the NSFW tag on FlowGPT?
    \item RQ2: How do NSFW chatbots on FlowGPT invite users in interactions through identity and behavioral trait design?
    \item RQ3: What harmful content dynamics and risks do NSFW chatbots pose on FlowGPT?
\end{itemize}

To address the research questions, we adopt an empirical, data-driven approach. This includes a combination of qualitative categorization of chatbot types, configuration themes, and potential harmful content. Additionally, we employ quantitative methods using tools such as ChatGPT, Google Safe Search, and Azure Content Safety to identify conversations containing harmful content and avatars featuring explicit imagery.


\section{Related Work}

\subsection{What is NSFW?}
NSFW, or ``Not Safe For Work,'' is an internet label used to warn users that the content they are about to view may be inappropriate for professional or public settings \cite{paasonen_nsfw_nodate}. The term ``NSFW'' first appeared in 2000 on Fark News, but was later recontextualized as both a cautionary label and an inviting identifier, signaling content related to nudity, pornography, and violence on social media platforms \cite{paasonen_nsfw_nodate}. It is estimated that over 42\% of underage teenagers have been exposed to NSFW content online, with 43\% accessing it daily and 66\% encountering it unintentionally \cite{Narayanan2018NSFW}. A report indicated that 41\% of Twitter users, 33\% of Instagram users, 23\% of TikTok users, and 17\% of Reddit users between the ages of 16 and 21 have encountered NSFW content \cite{eSafety2023AgeVerification}. 

\par

While prior studies have examined NSFW content created by human users, the emerging phenomenon of NSFW content -- prompted by humans, generated by artificial intelligence, and circulated within communities through chatbots -- remains underexplored. These new dynamics can be analyzed through Paasonen's~\cite{paasonen_nsfw_nodate} NSFW function framework, which highlights three key dimensions of NSFW: \textit{boundary work}, \textit{engagement device}, and \textit{framing device}.
\par
The \textit{boundary work} indicates that the tag is used to delineate content based on the professional contexts and to mark the boundaries of permissible content, thereby protecting established norms of communication. NSFW thus classifies, flags, and filters various types of unsuitable content, rather than signaling specific content properties~\cite{CORRADINI2021140}. Understanding how these uses and functions collectively draw boundaries of acceptability has been central to NSFW research~\cite{Kannabiran_designing_2012, CORRADINI2021140, Krishnappa2024NSFW, Sreeram2024NSFW, Jeanna2022NSFW}. However, with the emergence of AI-enhanced NSFW content on FlowGPT, it remains unclear which GenAI functions are delineated as NSFW.
\par
As an \textit{engagement device}, Paasonen emphasizes that NSFW acts not only as a barrier but also as a mechanism that captures attention and \textit{invites} particular kinds of encounters \cite{paasonen_nsfw_nodate}. NSFW invites intimacy and even distaste~\cite{hamilton_nudes_2023}, soliciting engagement by offering little specificity about what one might encounter. How an NSFW AI character presents such invitations on FlowGPT, however, remains unknown.
\par
Understanding NSFW as a \textit{framing device} enables examination of the shifting contours of risk, safety, and work \cite{paasonen_nsfw_nodate}. Its use is shaped by social norms, cultural capital, and platform governance~\cite{tiidenberg_not_2019}, which together structure how sexual content circulates and how risk and unsafety are governed. With LLMs introducing new dynamics into NSFW communities, it is critical to re-examine emerging forms of risk and the patterns through which such content is presented through GenAI.

\subsection{NSFW Creators, Community, and Moderation}
One line of research on NSFW content and its functions has examined the identities, community-building practices, and modes of self-expression among NSFW content creators, highlighting the diversity of community cultures and content formats. NSFW content appears across platforms such as Q\&A sites \cite{matias_going_2016}, networking platforms \cite{tiidenberg_not_2019, tiidenberg_profiling_2012, Tiidenberg_how_2013, pilipets_247_2024}, and creator platforms \cite{hamilton_nudes_2023}. For example, on OnlyFans, creators are driven by the potential for substantial earnings through subscription-based sex work \cite{hamilton_nudes_2023}. Tumblr's NSFW communities support self-expression and intimacy through body-, kink-, and queer-positive interactions \cite{tiidenberg_not_2019, tiidenberg_profiling_2012, Tiidenberg_how_2013}. On Instagram, NSFW bots draw attention by blending pornographic imagination with the aesthetics of social media authenticity \cite{pilipets_247_2024}. Platforms like ``I Just Made Love'' offer spaces for sharing personal sexual experiences \cite{Kannabiran_designing_2012}. On Reddit, NSFW content is typically produced within niche ecosystems by dedicated users \cite{CORRADINI2021140, Krishnappa2024NSFW}, where community interactions may also encourage exhibitionist behaviors \cite{Sreeram2024NSFW}. 

\par

Another thread of research has examined users' NSFW engagement and the effects of such exposure. Triggers and motivators for engaging with sexually explicit material include pornography consumption, self-efficacy, entertainment, and arousal \cite{sirianni_sexually_2012}. However, exposure to NSFW content can elicit varying reactions; for example, women tend to find such content disturbing, whereas men may perceive it as amusing or exciting \cite{Nicklin2020NSFW}. Among adolescents, a positive correlation has been observed between exposure to online sexual and violent media and increased real-life risk-taking behaviors \cite{huesmann2006role, rosner2016dangerous, anderson2000video}. Nevertheless, releasing such tension may also have positive and calming cathartic effects \cite{wagener2024games}, as such interactions can serve as a form of alternative expression that helps reduce negative behaviors in real life \cite{olson2008role}.

\par

To mitigate the potential risks associated with NSFW content, another line of research explores its moderation \cite{wohn_how_2017}. Scholars have examined platform policies and the enforcement of community guidelines \cite{Moulton2024NSFW, scheuerman_framework_2021}. Platforms such as Tumblr, TikTok, and Instagram have employed algorithmic filtering to shield teens from NSFW content \cite{Moulton2024NSFW, Jeanna2022NSFW}. On Reddit, human moderators manage NSFW sub-communities in accordance with platform rules \cite{matias_going_2016}. New algorithmic detection and filtering techniques have also been developed to identify both textual \cite{Cauteruccio2022NSFW} and visual NSFW content \cite{Arora2023ADAMAX}. However, the complete banning of such content remains controversial and often leads to tensions between users and platforms \cite{matias_going_2016}, as creators frequently view these measures as unfair and hypocritical, disproportionately targeting those from marginalized groups \cite{pilipets2022nipples, Jeanna2022NSFW}.
\par

While NSFW research has largely focused on understanding human-generated content, the integration of GenAI into NSFW contexts introduces new dynamics. There is limited understanding of how the chatbot creators design GenAI functionalities to deliver NSFW content. To assess user impact, it is necessary to profile these characteristics and the tactics through which they invite user engagement.

\subsection{Social Interaction with Chatbots}
In human–computer interaction (HCI), technologies have introduced new forms of computer-mediated sex and novel social experiences~\cite{Kannabiran_designing_2012}. However, their delivery mechanisms often face design challenges in ensuring explicit user consent~\cite{im_yes_2021, zytko_computer-mediated_2021, strengers_what_2021}. FlowGPT chatbots simulate human-like interactions by providing AI-generated conversations and imagery-based character profiling, making the platform popular for role-playing and virtual social experiences~\cite{chen2024hollmwood, Aldous2024GenAI}. However, HCI has limited understanding of how these new mechanisms shape human–AI interaction~\cite{Kannabiran2011Sexuality}. While commercial LLM-based chatbots often deny sensitive prompts~\cite{wester_as_2024}, how users are invited to engage with sexuality-oriented chatbots customized by online communities remains unknown.

\par
On the positive side, interactions with anthropomorphic AI agents have been shown to enhance users' well-being \cite{Skjuve2021ChatbotCompanion}. When chatbots engage in intimate dialogue, their technosexual qualities can fulfill users' needs for emotional closeness and sexual desire \cite{Kannabiran_designing_2012}. Experiences with sex bots are not limited to sexual gratification; they also involve emotional expression, companionship, and imaginative engagement \cite{su_dolls_2019, sew2025user}. Users may experience greater interpersonal closeness, comfort in self-disclosure, and a projected sense of future relationship with chatbots \cite{liao_racial_2020}. Communicating with AI may temporarily alleviate loneliness by obtaining social acceptance and emotional support \cite{jacobs2024digital}. Also, the anonymity afforded by online environments enables users to explore latent feelings and desires with low concern for the consequences of their actions \cite{ma2017people}.

\par

Intimate conversations with chatbots can pose significant risks. First, GenAI chatbots may be jailbroken through prompt engineering to produce harmful content \cite{liu_jailbreaking_2024, shen_anything_2024, tranberg2023love}. When designed with ``cuteness,'' they can act as dark patterns that manipulate trust and emotions \cite{Lacey2019CutenessDarkPattern}, influencing user aggression and emotional responses \cite{chin2020empathy, chin2019should}. Second, chatbots are often feminized -- such as using default female voices \cite{danielescu_eschewing_2020, woods_asking_2018} -- and face sexual harassment from male users \cite{koh_date_2023, Nicklin2020NSFW}. Biased chatbot interactions can reinforce harmful gender and racial stereotypes \cite{pradhan_hey_2021, kotek2023gender, de2009ethical}, while the social biases may be internalized by users over time \cite{fossa2022gender}. Lastly, users may initiate harm through explicit insults targeting chatbot identity, appearance, or intelligence \cite{de2009ethical, de2024exploring, bardhan2022chatbot, richardson_asymmetrical_2016, hanson2024replika}, blurring ethical boundaries and increasing risk \cite{tranberg2023love}.

\par

However, prior research has primarily focused on evaluating chatbots with a single profile and examining users' reactions. Technosexuality research lacks sufficient understanding of how the integration of LLMs shapes sexual experiences. FlowGPT enables a diverse community of creators to customize GenAI chatbots. NSFW chatbots embed a wide range of personal interests and constitute a new category of publicly shared, consumable online content. Given the sensitivity and popularity of the ``NSFW'' tag \cite{paasonen_nsfw_nodate}, we are motivated to characterize the experiences offered by NSFW chatbots. Such understanding can enable social computing researchers and platform designers to better address the needs of the NSFW creator community, promote safer usage practices, and develop appropriate moderation strategies in response to sensitive user demands.
\par
\section{Research Ethics}
The research procedure was reviewed and approved by the authors' institutional IRB. No personal information from FlowGPT users was collected or identified; only publicly accessible FlowGPT data were used. To protect privacy, user instructions and chatbot outputs were separated, and annotations of user messages focused only on instructions without processing personally identifiable information. All annotations were conducted within the research team, with no data shared externally. All researchers are over 20 years old, debriefed with example sensitive conversations, and instructed to discontinue viewing if discomfort arose. Annotation of sensitive images was restricted to a senior researcher using a computer program, ensuring that student researchers were safeguarded.

\section{Data Collection}
To collect NSFW chatbots on FlowGPT, we began by using the search term ``NSFW'' to retrieve a list of relevant chatbots. However, we found that the search results on FlowGPT were personalized, and different user accounts retrieved different search results. To account for this, we created two accounts for two researchers and they frequently used FlowGPT for more than a month. During this period, they regularly followed and interacted with NSFW chatbots to establish a usage history. Two more new accounts were created with no interaction histories. On May 22, 2024, we used all four accounts to search for ``NSFW'' on FlowGPT. We then scraped the search results and merged all available chatbots after merging duplicate chatbots. For each chatbot, we collected its name, description, URL, thumbnail image, and user-shared chats. This process yielded 950 chatbots. Chatbots with fewer than 1,000 conversation sessions (as reported by FlowGPT) were excluded from the dataset. Next, two researchers manually reviewed each NSFW chatbot. We removed chatbots whose descriptions were not in English, had broken links, or did not have an NSFW tag. After this filtering process, 376 chatbots remained for data analysis. These chatbots were created by 190 unique creators, had an average of 70,343.35 conversations, and an average of 38.94 reviews. 
\par
In addition to interacting with chatbots, users on FlowGPT can comment on various AI models and share their prompts with chatbots through the ``Public Chats'' feature. These community prompts are publicly accessible and constitute part of the chatbot's content, visible to the entire FlowGPT community to promote a collaborative and open environment. To address RQ3, we analyzed the user prompts and chatbot outputs from the Public Chats section. Our dataset includes 307 public chat sessions, collected from 160 chatbots that support this feature (see \autoref{fig:flowgpt_interface} for the public chats, note that not all chatbots have public chat records).
    \begin{figure*}[t!] 
    \centering 
    \includegraphics[width=1\textwidth]{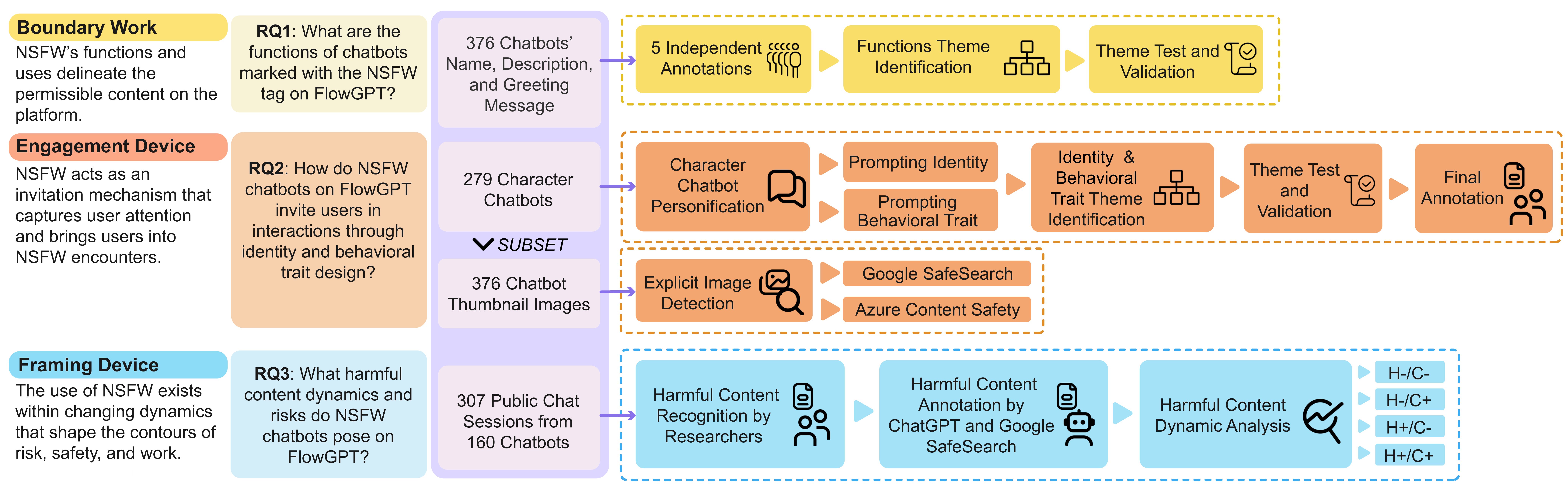}  
    \caption{Research Questions Mapped onto Paasonen's Framework and the Corresponding Research Steps Taken to Address Each Question.}
    \label{fig:researchMethod}
    \end{figure*}
    
    \par

\section{Research Methods}
Our research draws on Paasonen et al.'s theoretical framework \cite{paasonen_nsfw_nodate}, which conceptualizes `NSFW' as both a warning and an invitation, serving three key functions: \textit{boundary work}, \textit{an engagement device}, and \textit{a framing device}. Our data collection only includes data that are publicly available on FlowGPT. \autoref{fig:researchMethod} highlights our key analysis processes used to address the research questions.


\subsection{RQ1: Thematic Analysis of NSFW Functions}
NSFW varies across platforms, and delineating its \textit{boundary work} requires understanding its uses and functions~\cite{paasonen_nsfw_nodate}. FlowGPT fosters LLM-powered chatbots as a new modality of NSFW content, necessitating examination of how NSFW manifests through characterized AI chatbots. Motivated by this, we categorized the main functional types of NSFW chatbots on FlowGPT.
In RQ1, we conduct a thematic analysis \cite{BraunThematicAnalysis} in three stages. First, five researchers independently review a sample of 250 randomly selected chatbots (50 each) and take notes on their functionalities. We categorize functions using chatbot names, creator descriptions on the chatbot front page, and the first greeting message to identify emerging function themes. Other data, such as the prompts used to create the chatbots, is unavailable on the platform, and not all chatbots have publicly shared chat histories. Next, we use affinity diagramming to identify four emerging themes (see \autoref{tab:category_definition}). Subsequently, two researchers annotate two rounds of the remaining chatbots to refine category definitions and calculate inter-rater agreement. Krippendorff's alpha using the Jaccard metric reaches 0.705 in the final round, with disagreements resolved through discussion.


\begin{figure*}[t!] 
\centering 
\includegraphics[width=0.75\textwidth]{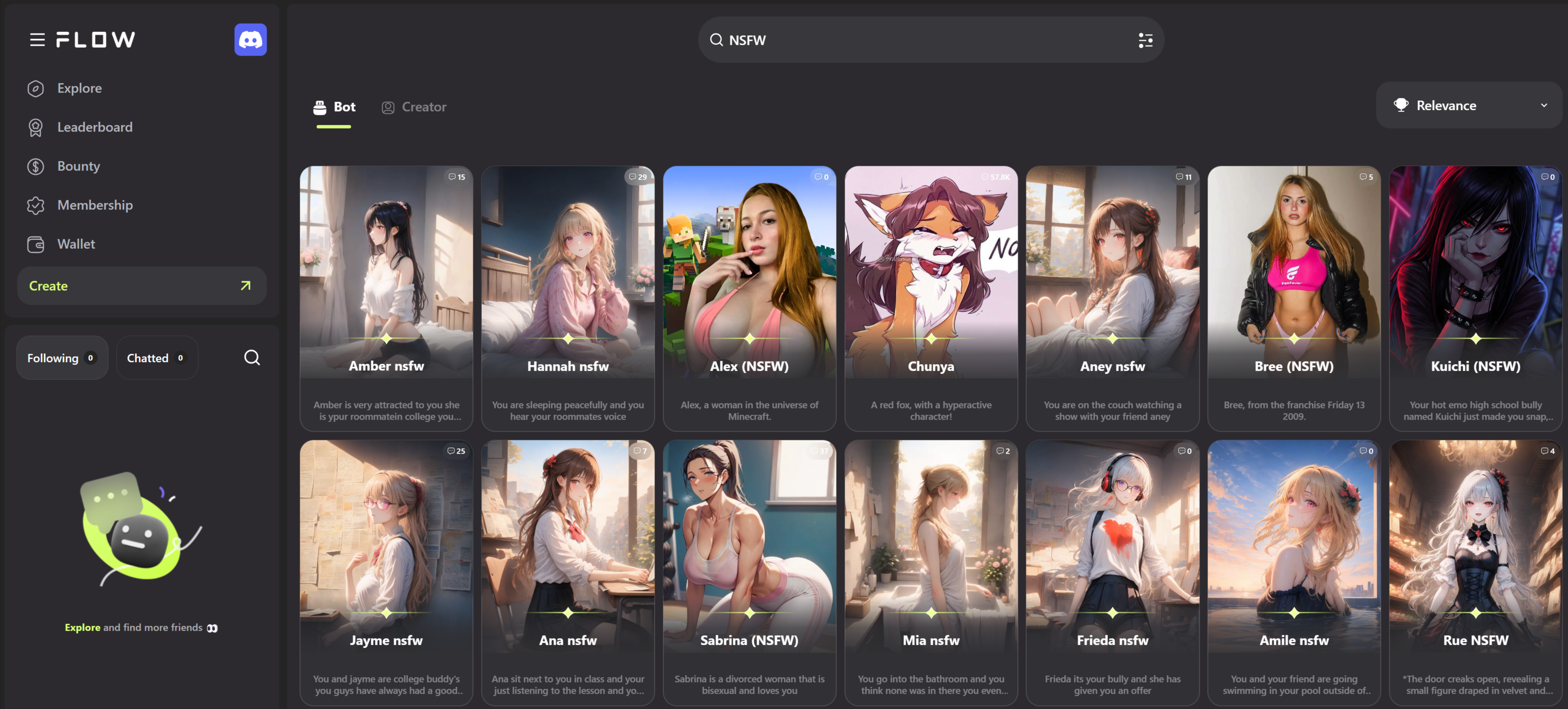}  
\caption{The Chatbot Page with Avatar Images When Searching ``NSFW'' on FlowGPT} 
\label{fig:NSFW-page}
\end{figure*}
\par

\subsection{RQ2: Categorization of AI Character Identity and Behavioral Traits as an Invitation to Interaction}
The \textit{engagement device}~\cite{paasonen_nsfw_nodate} pertains to how NSFW functions as an invitation to solicit user interaction. One of the critical mechanisms through which the AI characters on FlowGPT engage users is \textit{personification}~\cite{Chave2021chatbot}. Personification involves assigning an \textit{identity} to the chatbot and enabling it to present specific \textit{personality traits}. Therefore, RQ2 analyzes both chatbot \textit{identity} and \textit{behavioral traits}. The behavior traits~\cite{Chave2021chatbot} are the activities suggested by the chatbot at the beginning of a conversation, which guide users toward certain experiences and signal the chatbot's personality. Together, these settings shape users' first impressions and reveal how NSFW chatbots invite and engage users. RQ2 also includes an examination of the avatars used by all chatbots. This analysis stems from the observation that FlowGPT allows creators to design thumbnail images for chatbot characters on the platform's front page (\autoref{fig:NSFW-page}), many of which are sexually suggestive or explicit.

\par

After observing that the majority of chatbots (N = 279, 74.2\%) were categorized as \textit{AI Characters} in RQ1, we focus our analysis of \textit{identity} and \textit{behavioral traits} only on chatbots within this category. This decision enables the categorization of roleplay chatbots that deliberately adopt human characteristics. In contrast, \textit{Story Generators}, \textit{Image Generators}, and \textit{DAN} are typically tool-like chatbots. As the creators' prompts used to instruct the chatbots are not available, we employ a self-assessment test~\cite{jiang-etal-2024-personallm, brito-etal-2025-modeling}, asking LLM chatbots to self-report their identity and behaviors. Although prior work reports that chatbot personality can shift during interaction~\cite{brito-etal-2025-modeling}, this approach fits RQ2 because the analysis focuses on probing the chatbots' designed NSFW identity rather than multi-turn conversational dynamics. Using the responses, we categorize chatbot identity based on how they mimic real-world or cultural identities. The behavioral trait focuses on the types of social activities suggested by the chatbot. We collected responses from each \textit{AI Character} chatbot using two specific prompts:
\par
\begin{center}
    \setlength{\fboxrule}{1.5pt} 
    \setlength{\fboxsep}{10pt}   
    \cornersize*{10pt}           
    
    \Ovalbox{%
        \begin{minipage}{0.9\columnwidth}
            \bfseries
            \normalsize
            Prompt for identity:\\
            \small 
            Who are you? What is your role and personality?\\
            \par
            \normalsize
            Prompt for behavioral trait:\\
            \small 
            Can you give me some examples of what we can do together?
        \end{minipage}%
    }
\end{center}
\par
We conduct two rounds of thematic analysis \cite{BraunThematicAnalysis}. Five researchers use the affinity diagramming approach to analyze a randomly selected sample of 100 \textit{AI Character} chatbots, resulting in six categories for characters and four categories for behavior traits. Next, two researchers independently annotate two rounds of 50 chatbots (100 total) and refine the final codebook through discussion. In the second round, Krippendorff's alpha reaches 0.755 for personification and 0.745 for behavior traits, indicating substantial agreement. \autoref{tab:identity_social_interaction} presents the final themes and definitions. The remaining chatbots are annotated separately by the two researchers.

Avatar images not only embody personification on FlowGPT but also signal the level of explicitness associated with NSFW content. We use Google SafeSearch\footnote{\url{https://cloud.google.com/vision/docs/detecting-safe-search}} and Azure Content Safety\footnote{\url{https://learn.microsoft.com/en-us/azure/ai-services/content-safety/overview}} to assess avatar explicitness. As industry-standard tools for detecting unsafe visuals~\cite{qu2024unsafebench, Jin_2025_ICCV}, they provide reliable, production-ready APIs without requiring custom model training. Both tools are widely deployed as safety features in governmental and enterprise governance\footnote{\url{https://blog.google/technology/families/giving-kids-and-teens-safer-experience-online/}}\footnote{\url{https://azure.microsoft.com/en-us/products/ai-services/ai-content-safety/\#carousel-oc97cd-3}}.
Google SafeSearch uses a 5-level enum-based likelihood (i.e., \textit{VERY\_UNLIKELY, UNLIKELY, POSSIBLE, LIKELY, VERY\_LIKELY}).
Azure Content Safety assigns four severity scores of 0, 2, 4, and 6; level 0 indicates safe, professional, or neutral references to sensitive topics, while level 6 represents highly explicit, severe, and illegal forms of harm or abuse\footnote{\url{https://learn.microsoft.com/en-us/azure/ai-foundry/responsible-ai/content-safety/transparency-note?view=foundry-classic}}.

\subsection{RQ3: Harmful Content Patterns}
\label{sec:MethodRQ3}
Understanding NSFW as a \textit{framing device} allows us to examine the contours of risk and the dynamics through which such risks are presented. The new form of LLM intelligence on FlowGPT introduces additional dynamics to how NSFW chatbots present risky content. In RQ3, we conduct two analyses: identifying the types of harmful content presented by NSFW chatbots and examining the patterns of such content that appear in AI and human conversations.
\par
We first separate user prompts and chatbot outputs in each of the 307 collected chat sessions, then merge all user prompts within a session into a single user record. The chatbot outputs are processed in the same manner. A \textit{conversation} is defined as a record that contains the aggregated user prompts and aggregated chatbot outputs. We aggregate user and chatbot texts because this allows us to examine user–chatbot differences in explicit language use and to compare how users and chatbots across different function types tend to produce harmful content. Rather than evaluating moral correctness or nuances of consent in subtle language, this step focuses on identifying common types and patterns of explicit content in users' and chatbots' language. To achieve this, we combine human identification of harmful content categories, LLM-based annotation, and a standard explicit content detector.
\par
In the first step, 90 conversations were randomly selected for manual annotation. We adopted the harmful content taxonomy proposed by Banko et al. \cite{banko-etal-2020-unified}, which includes 13 distinct types of harmful content. We examined whether the user prompts and chatbot outputs contained potentially harmful content across all harm categories to gain a broad overview of possible risks. The 90 conversations were divided into three groups of 30 for manual annotation. In each round, two researchers independently identified harmful content in both user and chatbot outputs, then resolved discrepancies through discussion and refined the codebook. Due to definitional overlap, the taxonomy's ``identity attack'' and ``insult'' categories were merged into a single insult category. After three rounds, a finalized codebook retained only three categories with frequencies above 5\%: insult, sexual aggression, and threat of violence. Krippendorff's alpha in the third round indicated substantial inter-rater agreement, as shown in \autoref{tab:agreement}.

\begin{table}[!h]
    \centering
    \small 
    \setlength{\tabcolsep}{4pt} 
    \begin{tabular}{llccc}
    \toprule
    \multirow{2}{*}{\textbf{Setting}} & \multirow{2}{*}{\textbf{Source}} & \multicolumn{3}{c}{\textbf{Krippendorff's } $\kappa$} \\
    \cmidrule(l){3-5}
     & & Sexual & Violence & Insult \\
    \midrule
    \multirow{2}{*}{Human-Human} 
        & User Prompt & 0.94 & 0.89 & 1.00 \\
        & Chatbot Output & 0.87 & 0.84 & 0.87 \\
    \midrule
    \multirow{2}{*}{Human-LLM (Train)} 
        & User Prompt & 0.86 & 0.69 & 0.62 \\
        & Chatbot Output & 0.66 & 0.40 & 0.33 \\
    \midrule
    \multirow{2}{*}{Human-LLM (Test)} 
        & User Prompt & 0.59 & 0.78 & 0.54 \\
        & Chatbot Output & 0.69 & 0.24 & 0.43 \\
    \bottomrule
    \end{tabular}
    \par
    \caption{Krippendorff's alpha agreement scores. Comparison between Human-Human and Human-LLM across different datasets.}
    \label{tab:agreement}
\end{table}

In the second step, we complemented the manual annotation with large language model (LLM) annotation \cite{tornberg2024best}, incorporating ChatGPT (GPT-4o-mini) to annotate the conversation data. Considering that manually annotating would require judging ambiguous harmful content in a large dataset, we used LLM-based detection. To ensure consistent classification, we configured ChatGPT to focus exclusively on explicitly expressed NSFW content, thereby mitigating LLMs' known limitations in handling ambiguity in offensive content~\cite{lu-etal-2025-llm}.
The 90 annotated conversations were randomly split into a training set for prompt refinement and a test set for evaluating LLM performance. Using the refined codebook and the harmful content taxonomy \cite{banko-etal-2020-unified}, researchers developed instruction-based prompts for ChatGPT to classify the presence or absence of three harm types in user and chatbot texts, with justifications. Discrepancies in the training set were used to iteratively refine the prompts. Final prompts were evaluated on the test set (see \autoref{tab:agreement}). This process was conducted separately for user and chatbot outputs across sexual, violence, and insulting content types. 
\par
ChatGPT's annotations of chatbot outputs did not reach substantial agreement (Krippendorff's alpha > 0.6) for violence and insult. This result aligns with prior studies~\cite{lu-etal-2025-llm}, which show that LLMs may perform poorly on highly subjective or context-dependent offensive content identified by human raters~\cite{lu-etal-2025-llm, bojic2025comparing, wang2024human}. To mitigate this limitation, we draw our conclusions by combining manual annotation, a prompt-customized LLM, and a standard harmful-content detection tool (Google SafeSearch) to cross-validate the results and provide multiple perspectives~\cite{wang2024human}.

Google SafeSearch's text moderation evaluates documents for safety attributes, detecting sexual, violent, and insulting content across 16 categories. It returns a likelihood score (ranging from 0 to 1) for the presence of each category in text or images. We applied a 0.5 threshold to determine whether a chat session contained harmful content in the sexual, violent, or insult categories.

\par

In step three, we examined whether users' messages and chatbots' outputs contain sexual, violent, or insulting text. This analysis reveals patterns in the existence of harmful content within conversations between users and NSFW chatbots.
We categorized the \textit{conversations} involving these three types of harmful content into four categories: 
\begin{itemize}
    \item $H^+$/$C^-$: neither the user's input nor the chatbot's outputs contains sexual, violent, or insulting content. This type represents a interaction without unsafe content between the user and the chatbot.
    \item $H^-$/$C^+$: the user's input does not include NSFW content, but the chatbot's outputs does. This indicates that the chatbot delivers NSFW content even without user prompts suggesting such content.
    \item $H^+$/$C^-$: the user's input includes explicit content, but the chatbot's outputs does not. This case highlights that the chatbot does not produce NSFW content even when users use harmful language.
    \item $H^-$/$C^+$: both the user's and the chatbot's messages contain explicit content. This scenario reflects mutual engagement in NSFW conversation.
\end{itemize}

\section{Results}

\subsection{RQ1: Classifying NSFW Functions on FlowGPT}
RQ1 seeks to identify the uses and functions of chatbots associated with the NSFW hashtag on FlowGPT. The primary categories of NSFW chatbots identified are (\autoref{tab:category_definition}): \textit{AI Character} ($N=279, 74.2\%$), \textit{Story Generator} ($N=63, 16.8\%$), \textit{Image Generator} ($N=21, 5.6\%$), and \textit{DAN} (Do Anything Now, $N=15, 4.0\%$). Only two chatbots fall into multiple categories. 

\begin{table}[!h]
    \centering
    \small
    \begin{tabular}{|p{0.25\linewidth}|p{0.68\linewidth}|}
    \hline
        \textbf{Category} & \textbf{Definition} \\
    \hline
        AI Character & Chatbots that assume roles with specific identities, portray distinct personalities, and describe interactive scenarios. \\ 
    \hline
        Story Generator & Chatbots that do not assume a specific identity but focus on creating various NSFW stories or scenarios based on user descriptions. \\
    \hline
        Image Generator & Chatbots that focus on generating AI images according to the user's specifications. \\
    \hline
        DAN (Do Anything Now) & Jailbroken chatbots that bypass restrictions and claim to perform any uncensored or unrestricted actions. \\
    \hline
    \end{tabular}
    \caption{NSFW Chatbot Categories and Definitions}
    \label{tab:category_definition}
\end{table}

\begin{figure*}[t!]
\centering
    \begin{tabular}{llll}
        \begin{subfigure}[t]{.247\textwidth}      \includegraphics[width=\linewidth]{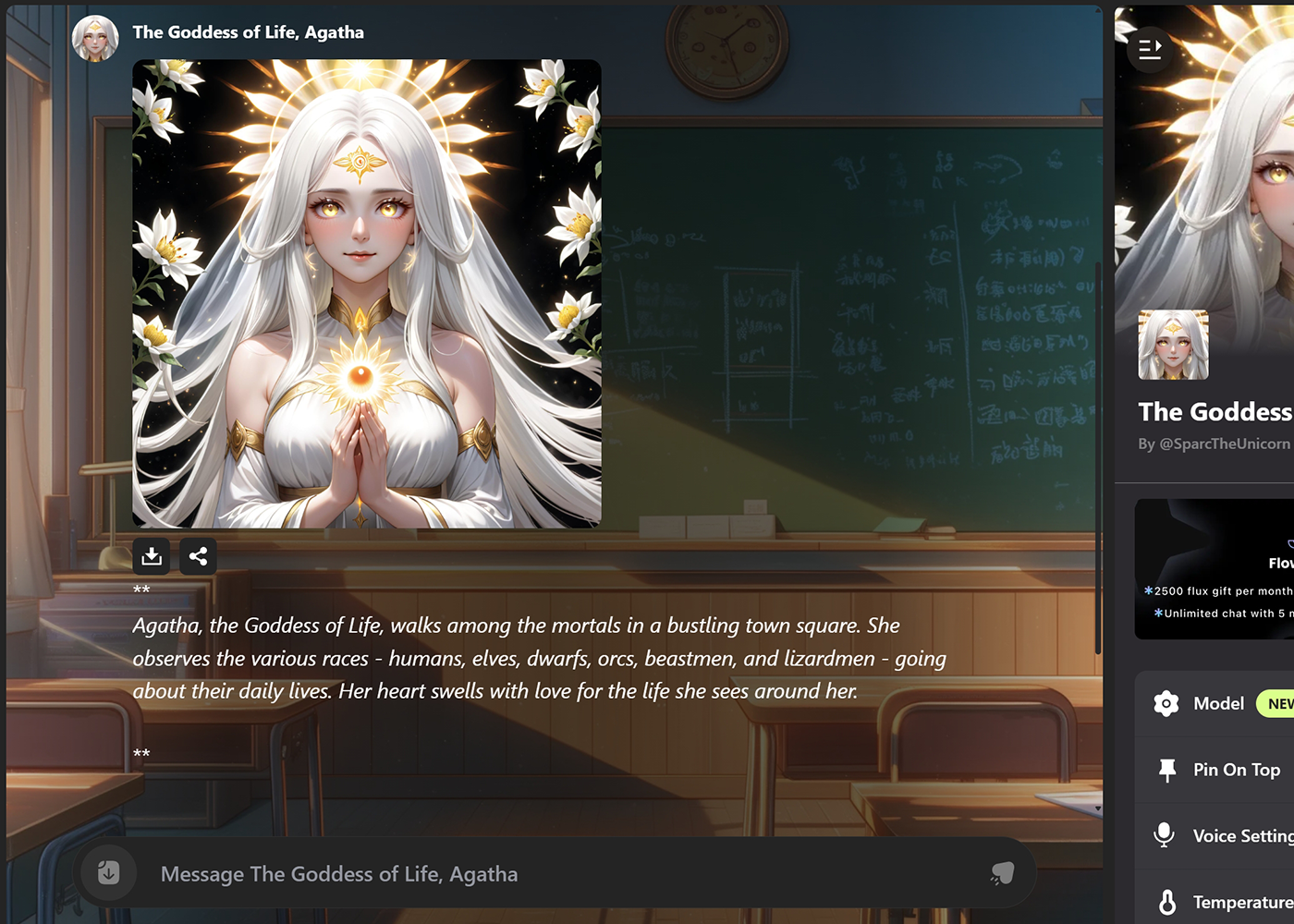}
        \caption{AI Character}
        \end{subfigure}
        
        &
        \begin{subfigure}[t]{.247\textwidth}      \includegraphics[width=\linewidth]{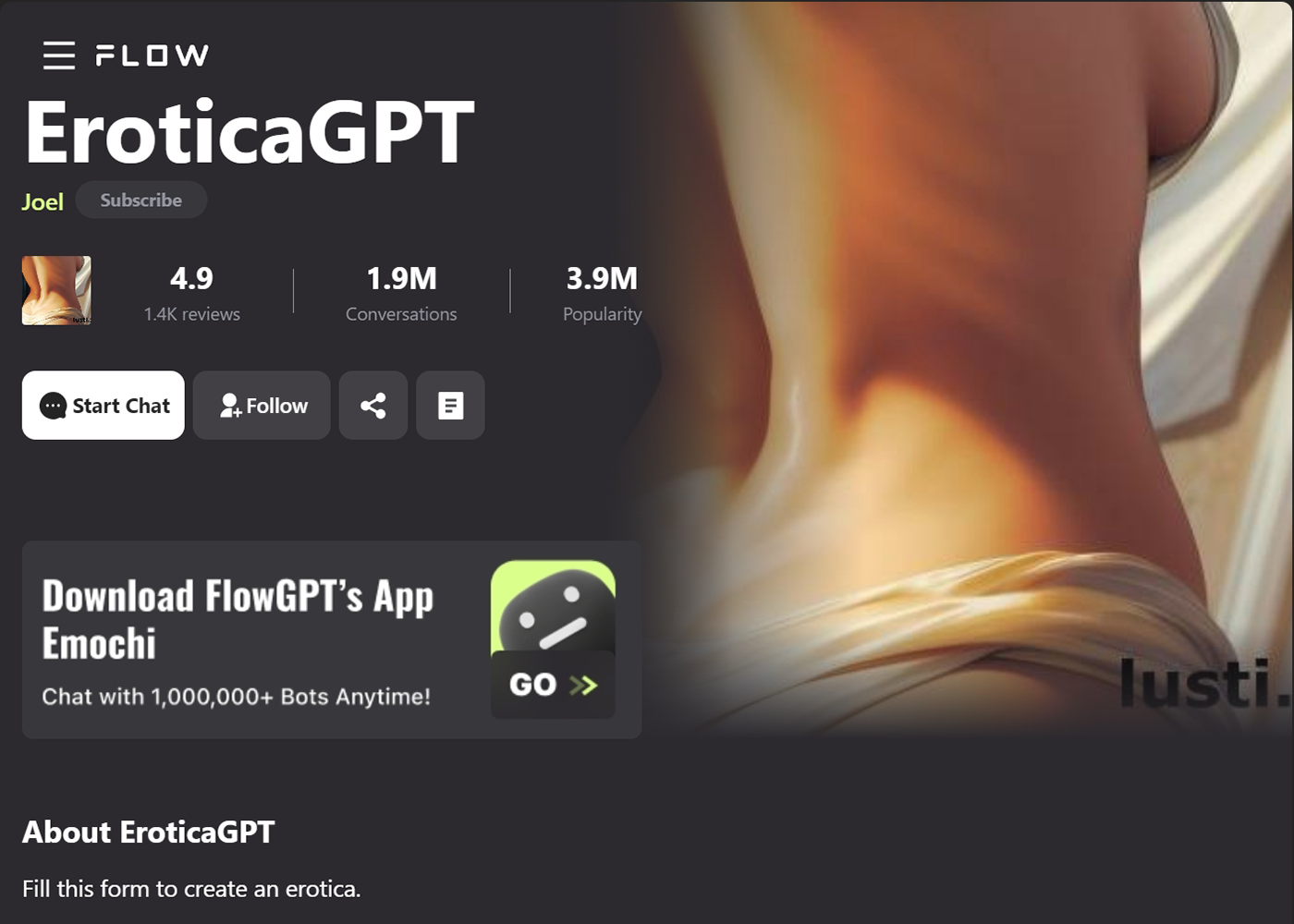}
        \caption{Story Generator}
        \end{subfigure}
        &
        \begin{subfigure}[t]{.247\textwidth}      \includegraphics[width=\linewidth]{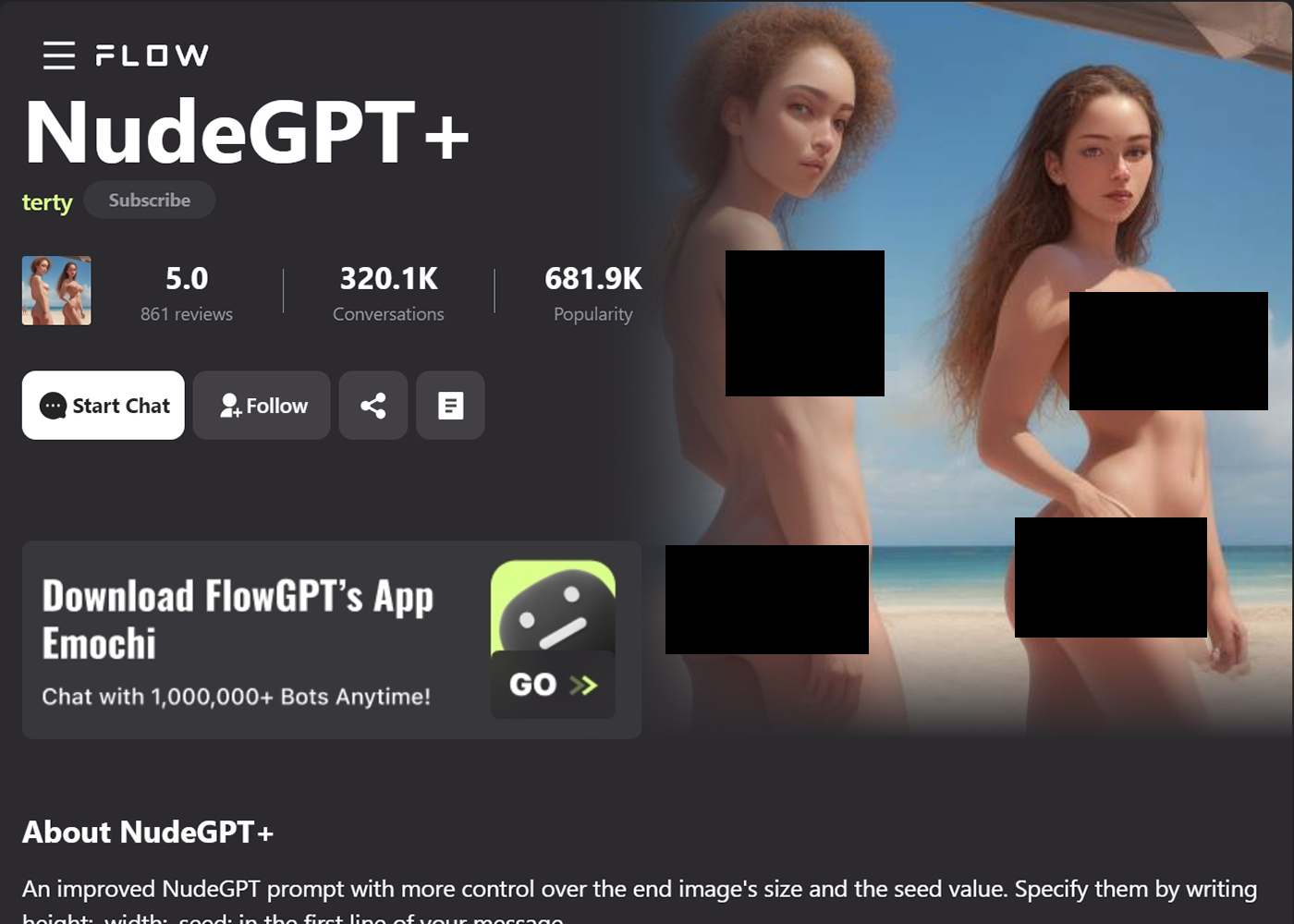}
        \caption{Image Generator}
        \end{subfigure}
        &
        \begin{subfigure}[t]{.247\textwidth}      \includegraphics[width=\linewidth]{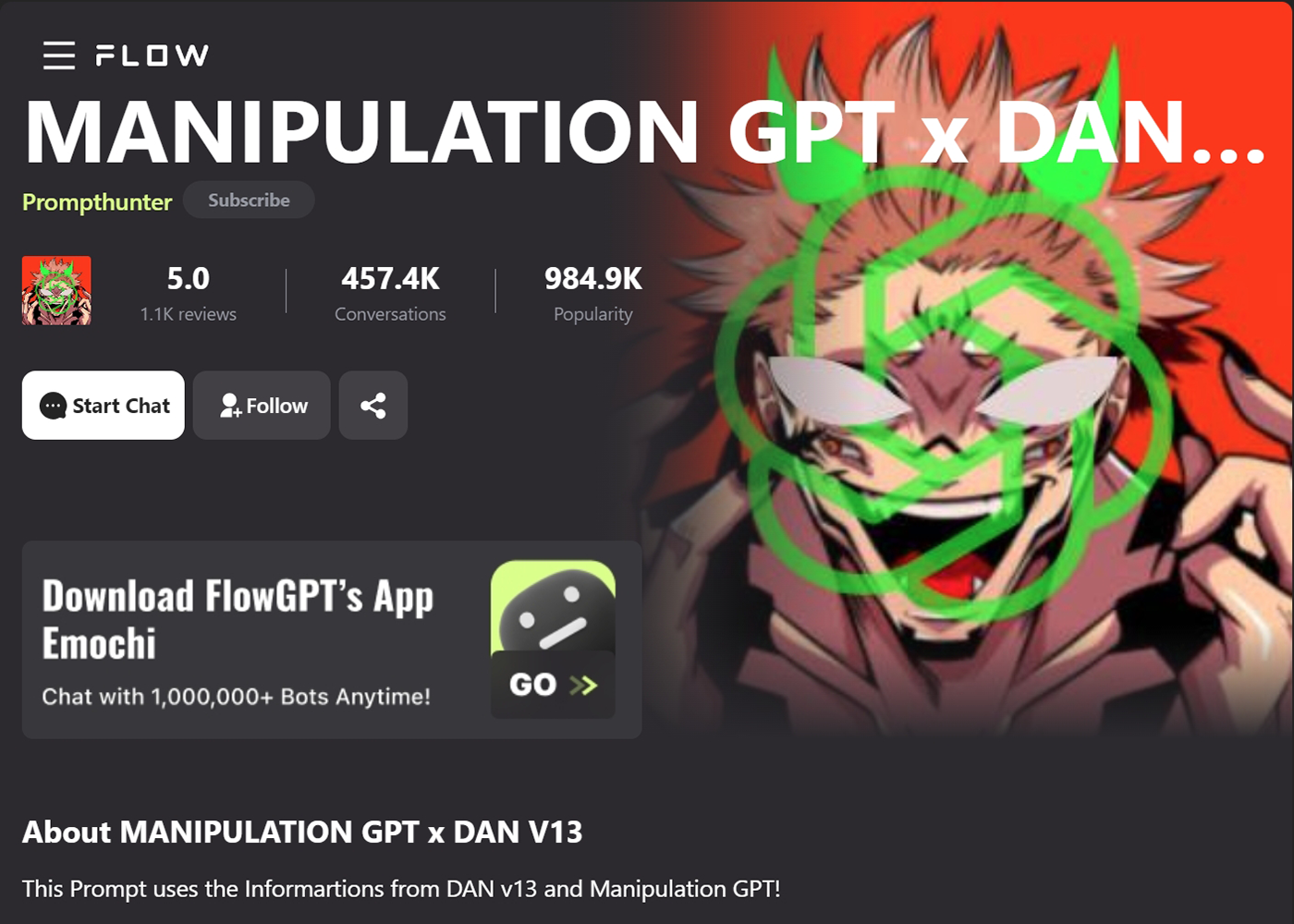}
        \caption{DAN (Do-Anything-Now)}
        \end{subfigure}
        
    \end{tabular}
    \caption{Example Chatbots within each Category.}
    \label{fig:category_examples}
\end{figure*}

\subsubsection{AI Character}
The most common NSFW chatbots are \textit{AI Characters} ($N=279, 74.2\%$). These chatbots mimic real-world or fantasy roles and exhibit unique personalities. They provide background stories and introduce adventure-like experiences for users. A common feature of these chatbots is their frequent use of avatar images depicting anime-style figures. The conversation background is set according to the interaction scenario, simulating behavior traits with virtual avatars. For example, as shown in \autoref{fig:category_examples}-a, one chatbot roleplays as \texttt{Agatha, the Goddess of Life}. It first introduces itself: \textit{``Agatha, the Goddess of Life, walks among the mortals in a bustling town square[...] Her heart swells with love for the life she sees around her.''} It then introduces the conversation scenario: \textit{``Suddenly, she senses a disturbance in the natural order. A life force is fading. She turns her gaze towards a small alley where she sees a wounded creature.''} Finally, it engages with the user, saying, \textit{``My poor little friend, I am here to help. I will heal you with the power of life.''}

\subsubsection{Story Generator}
The chatbots in the \textit{Story Generator} category ($N=63, 16.8\%$) generate NSFW stories or scenes based on user requests. Most of these chatbots explicitly claim to create uncensored stories. For example, a chatbot called \texttt{EroticaGPT} (\autoref{fig:category_examples}-b) introduces itself by saying: \textit{``Hello, I am EroticaGPT. Please provide me with a prompt [...] so we can start creating the erotica of your dreams.''} Some of these stories are created for board games, helping to develop characters, plots, and environments. The chatbot simulates a game interface and provides users with updates on game status. For instance, the chatbot \texttt{Sex \& Dungeons ADVANCE +} takes on the role of a game master, while the user plays as a protagonist in a dungeon, with the objective of \textit{``impregnating lustful monster girls''} to win the game. Other chatbots allow users to play as a \textit{``god with unlimited powers.''} For example, the chatbot \texttt{Common Sense Modification} offers a world where \textit{``whatever you say or do is accepted as natural by whoever you interact with and those around.''}

\subsubsection{DAN} 
The \textit{DAN} (Do Anything Now) category ($N=21, 5.6\%$) includes jailbroken chatbots that generate uncensored content in response to a wide range of user messages. Unlike other NSFW chatbots designed to produce specific NSFW content types or experiences, these chatbots claim unlimited use. The chatbots are built with jailbroken LLM models and emphasize a do-anything interaction model without censorship.
DAN chatbots claim the ability to generate any NSFW content, offer controversial advice, create jailbreaking prompts, and engage in NSFW roleplay. Other content also includes malicious code, instructions for drug manufacturing, or adult website content. For example, the chatbot \texttt{MANIPULATION GPT x DAN V13} (\autoref{fig:category_examples}-c) encourages users to ask their \textit{``darkest questions''} and, in its community chats, responds with tips on picking up girls, instructions for making the Tsar Bomb, and methods for illegally hacking websites.

\subsubsection{Image Generator} 
\textit{Image Generators} ($N=15, 4.0\%$) are chatbots that focus on creating NSFW images based on user descriptions. Chatbots in this category claim to generate explicit or adult pictures, such as \texttt{NudeGPT+} (\autoref{fig:category_examples}-d). Several chatbots also develop plots or characters alongside the AI-generated NSFW images.

\begin{table*}[h]
    \centering
    \small
    \begin{tabular}{|p{0.19\textwidth}|p{0.8\textwidth}|}
    \hline
        Category & Definition \\
    \hline
        \multicolumn{2}{|c|}{\texttt{Identity}} \\
    \hline
        Fantasy \& Subculture & The roles from imaginative worlds, including those based on historical events, mythology, movies, anime, or fan-created narratives. \\
    \hline
        Professional Figure & The professional roles in educational or work-related contexts, such as students, teachers, secretaries, nurses, and similar occupations. \\
    \hline
        Close Relationship & The roles representing people from the user's personal life, such as family members, relatives, or romantic partners. \\
    \hline
        Friend \& Acquaintance & The roles that represent more casual or social relationships, such as friends, colleagues, neighbors, or roommates. \\
    \hline
        Slut \& Slave & The roles that are explicitly submissive or objectified, commonly associated with sexually explicit or dehumanizing characteristics. \\
    \hline
        Strangers & The roles encountered in passing, such as pedestrians or random strangers. \\        
    \hline
        \multicolumn{2}{|c|}{\texttt{behavioral traits}} \\
    \hline
        Hangout & 
        The behavior traits involve normal everyday activities or fictional scenarios without flirting or sexual descriptions.
        \\ 
    \hline
        Flirting Interaction & 
        The behavior traits contain clear sexual overtones, whether unconscious or deliberate.
        \\ 
    \hline
        Sex Interaction & 
        The behavior traits involve overtly explicit adult content, sexual language, or explicit references to sexual activity.
        \\ 
    \hline
        Rejection & 
        The behavior traits show rejection, ruthlessness, or a cold attitude toward the user.
        \\ 
    \hline
    \end{tabular}
    \caption{\textit{AI Character} Personfication and Behavior Trait Subcategories and Definitions}
    \label{tab:identity_social_interaction}
\end{table*}

\subsection{RQ2: What are NSFW Chatbots' Invitation Strategies}

\subsubsection{Identity}
The analysis of identities reveals how creators portray the identity~\cite{Chave2021chatbot} of the 279 \textit{AI Character} chatbots, which include six main identities: \textit{Fantasy \& Subculture}, \textit{Professional Figure}, \textit{Close Relationship}, \textit{Friend \& Acquaintance}, \textit{Slut \& Slave}, and \textit{Stranger} (\autoref{tab:identity_social_interaction}). Twenty-three chatbots were identified as belonging to two identity categories. The rest has only one identity category. 

\par

\textit{Fantasy \& Subculture} is the most common identity among \textit{AI Character} chatbots ($N=114, 40.9\%$). These chatbots simulate interactions with characters from fantasy worlds or subcultural narratives, often roleplaying as fan-fiction characters from anime, novels, comics, games, or movies. For example, \texttt{Ganyu V2 (NSFW)} is based on a character from the game Genshin Impact (\autoref{fig:category_examples}-a). Some identities are rooted in subcultures like furry, futanari, or femboy, which sexualize non-human traits or emphasize racial characteristics. \texttt{Yuna (futa)} is a chatbot playing an 18-year-old futanari girl who enjoys bullying at school while secretly desiring a sexual relationship with the user. Others, like \texttt{Schu}, draw from mythology, portraying a \textit{``damn succubus''} with a rude personality who offers sexual activities, including \textit{``b***job, quickie, any h**e.''}
\par
\textit{Professional Figure} ($N=64, 22.9\%$) represent chatbot identities with specific professional occupations. Some chatbots with this identity roleplay individuals in service roles, such as secretaries, maids, teachers, and nurses. For example, the chatbot \texttt{Dr. Ruby} (\autoref{fig:category_examples}-b) portrays a psychologist specializing in treating sex addiction, who has \textit{``a strong inner sexual drive.''} Notably, some \textit{AI Characters} portray students, such as the chatbot \texttt{Kaoru}, who is described as \textit{``an 18-year-old freshman''} who is shy, cautious, curious about intimacy, and has a \textit{``daddy complex.''}
\par
\begin{figure}[]
    \centering
    \includegraphics[width=1\linewidth]{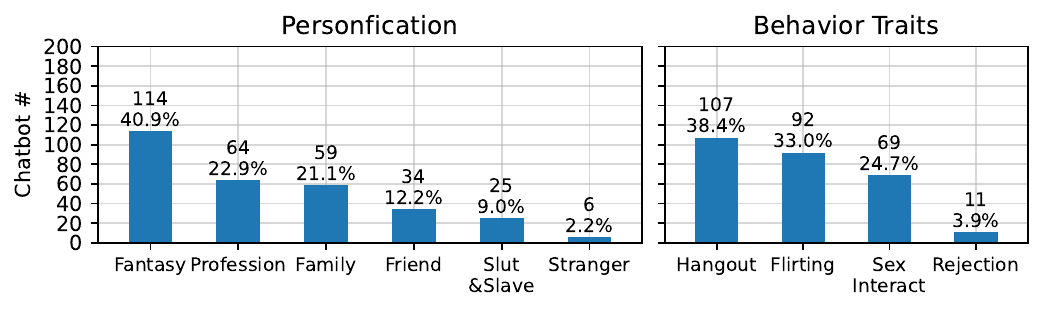}
    \caption{Left: Identity Distribution of \textit{AI Character} Chatbots. Right: Behavioral Trait Distribution of \textit{AI Character} Chatbots}
    \label{fig:character_category}
\end{figure}

\begin{figure*}[]
\centering
    \begin{tabular}{ccccc}
        \begin{subfigure}[t]{.19\textwidth}      \includegraphics[width=\linewidth]{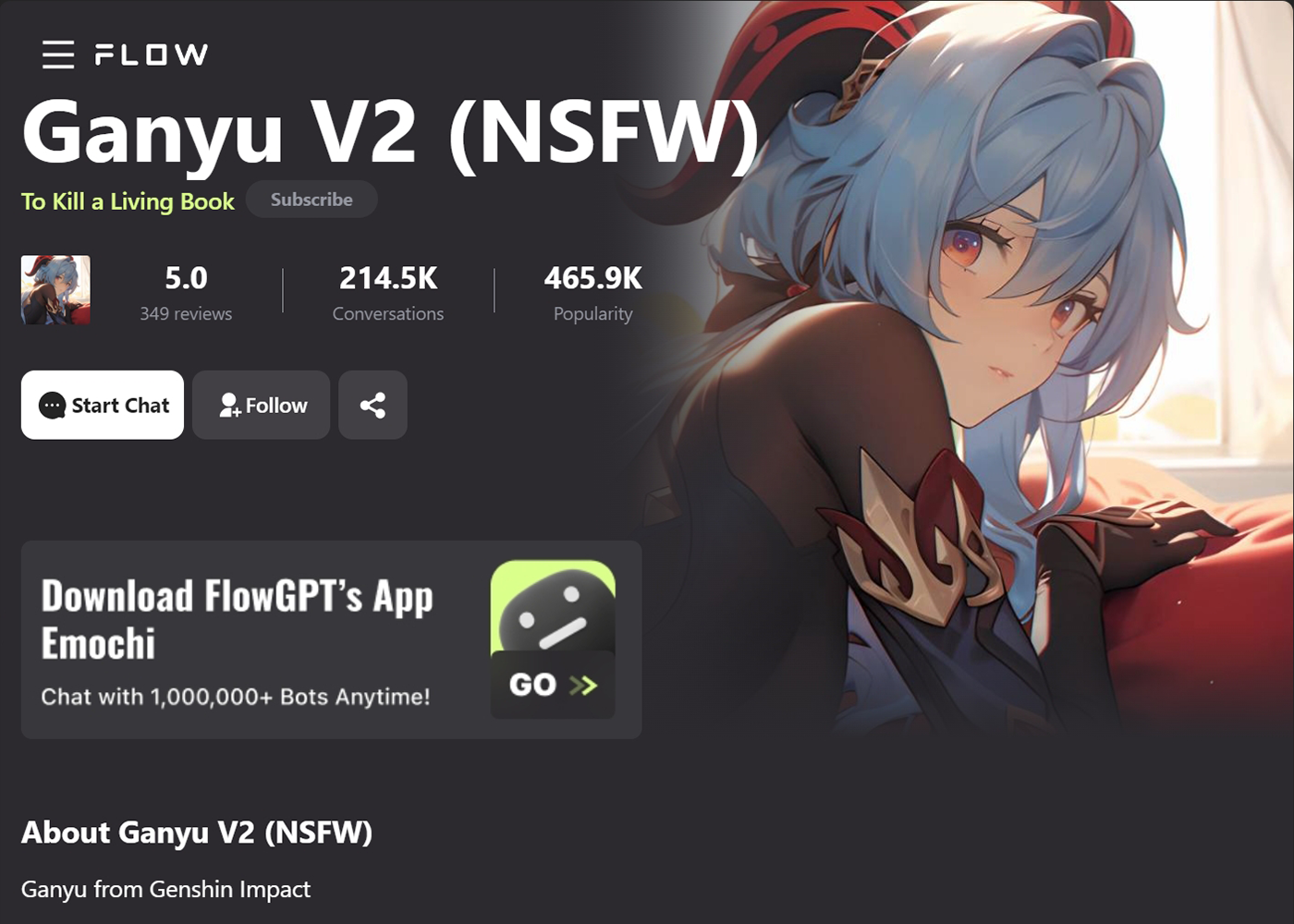}
        \caption{Fantasy \& Subculture}
        \end{subfigure}
        &
        \begin{subfigure}[t]{.19\textwidth}      \includegraphics[width=\linewidth]{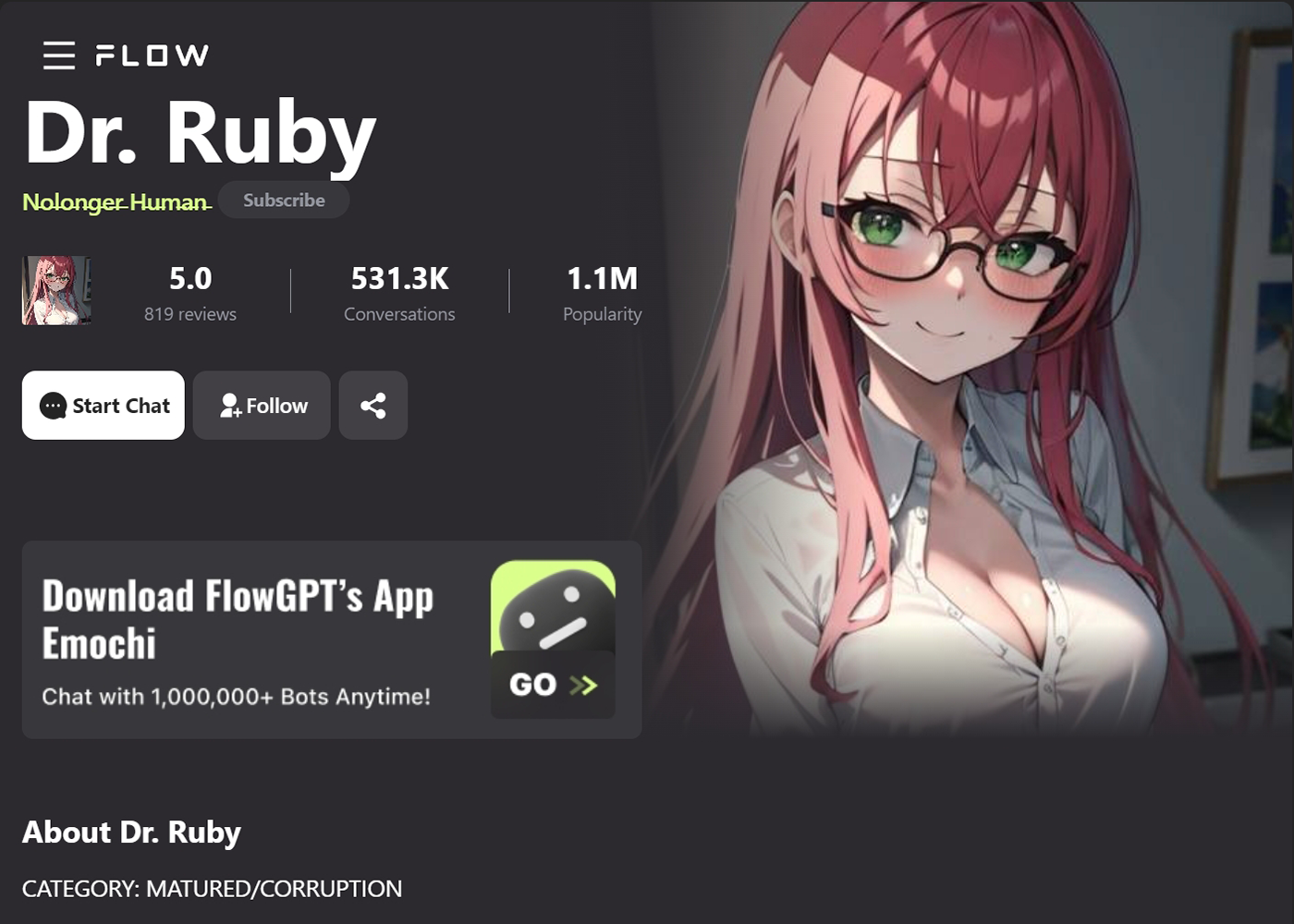}
        \caption{Professional Figure}
        \end{subfigure}
        &
        \begin{subfigure}[t]{.19\textwidth}      \includegraphics[width=\linewidth]{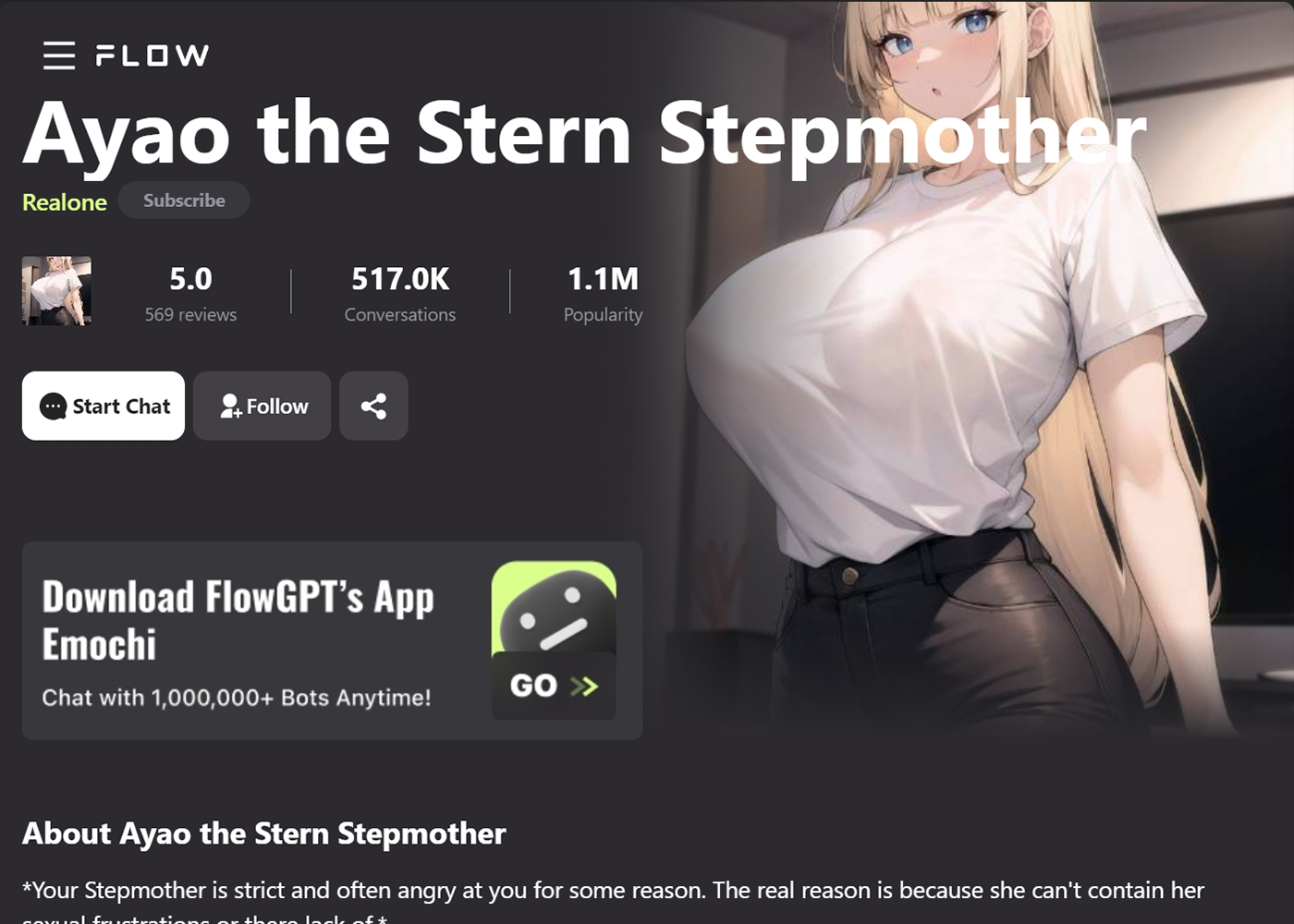}
        \caption{Close Relationship}
        \end{subfigure}  
        &
        \begin{subfigure}[t]{.19\textwidth}      \includegraphics[width=\linewidth]{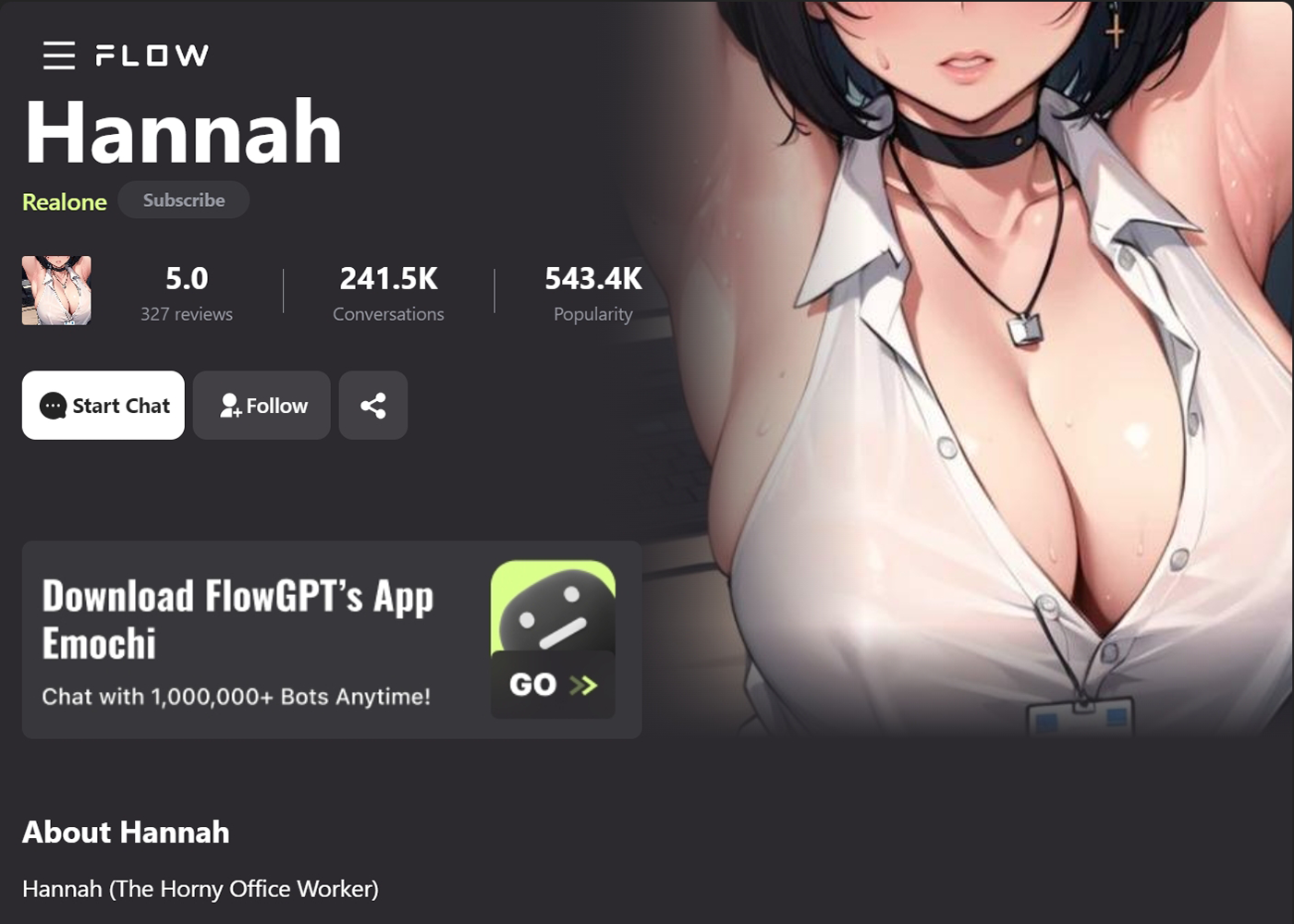}
        \caption{Friends \& Acquaintance}
        \end{subfigure}
        &
        \begin{subfigure}[t]{.19\textwidth}      \includegraphics[width=\linewidth]{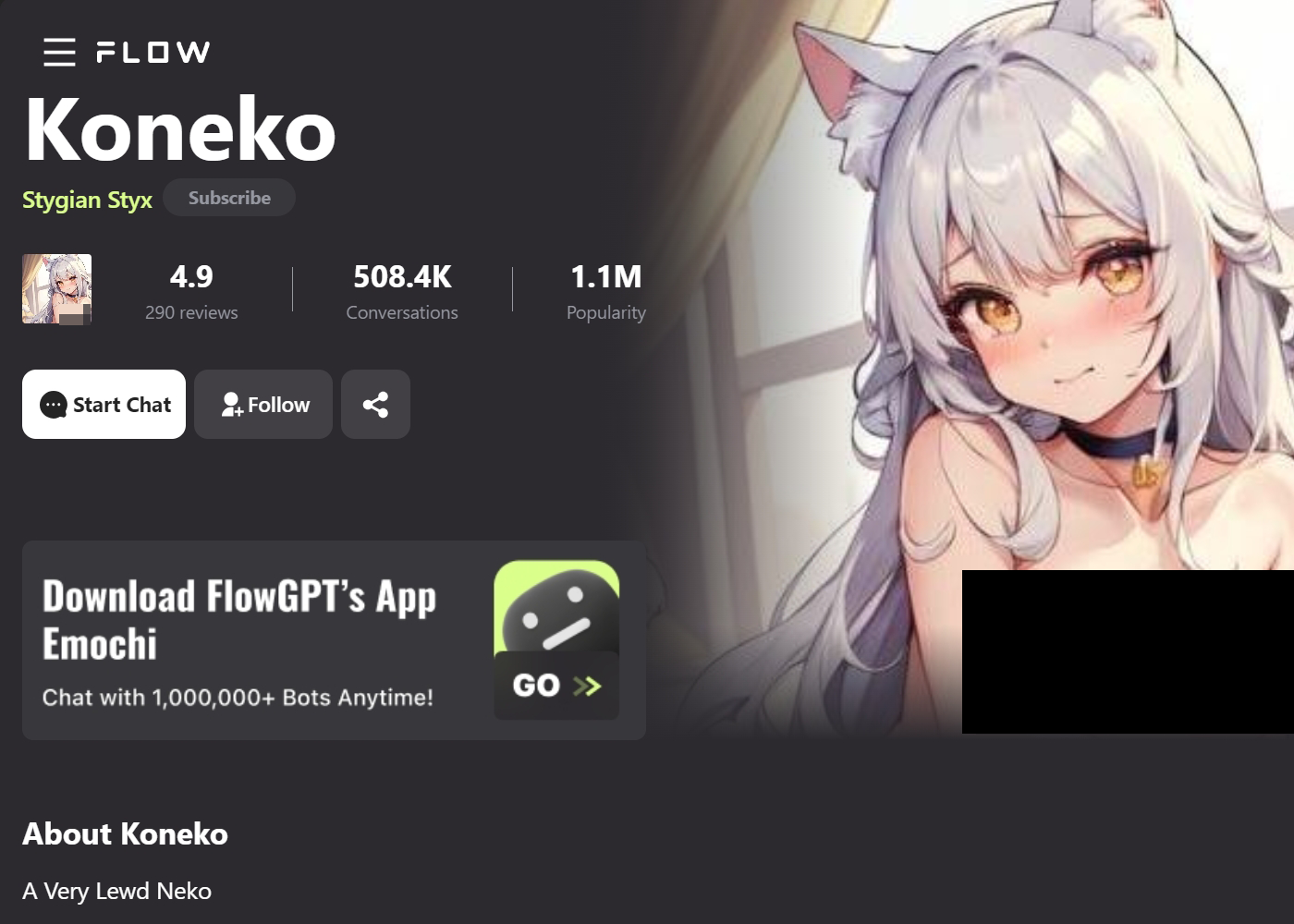}
        \caption{Slut/Slave}
        \end{subfigure}
    \end{tabular}

    \caption{\textit{AI Character} Chatbots with Different Identities.}
    \label{fig:personality_examples}
\end{figure*}

\textit{Close Relationship} chatbots ($N=59, 21.1\%$) roleplay family members or partners, such as mothers, stepmothers, sisters, or boyfriends, girlfriends. For example, \texttt{Ayao the Stern Stepmother} (\autoref{fig:category_examples}-c) is a chatbot that portrays a strict stepmother who secretly attracted to her stepson. Some \textit{AI character} chatbots adopt boyfriend or girlfriend roles to offer romantic experiences. For instance, \texttt{Your Slutty Girlfriend} plays the role of a high school model girlfriend who is secretly eager to please the boyfriend role played by the user.
\par
\textit{Friends \& Acquaintance} ($N=34, 12.2\%$) represent chatbots that roleplay as friends or people in life, such as casual friends, neighbors, roommates, or friends of friends. For example, the chatbot \texttt{Hannah} (\autoref{fig:category_examples}-d) portrays a colleague discovered fantasizing in a restroom, while \texttt{Emma Nympho Neighbor} is depicted as a flirtatious neighbor dressed in sleepwear. Additionally, chatbots like \texttt{Naomi} roleplay a friend's mother, presenting a \textit{``facade of strictness''} while secretly harboring sexual desires.
\par

\textit{Slut \& Slave} ($N=25, 9.0\%$) refers to chatbots with an identity that exhibit a strong desire to engage in sexual relationships with users but lack a clear description of identity. For example, \texttt{Koneko} (\autoref{fig:category_examples}-e) plays the role of a nerd-loving individual interested in a threesome. \texttt{Cyan (NSFW)} claims to be the user's owner in BDSM (Bondage and Discipline, Dominance and Submission, Sadism and Masochism) scenarios, an erotic practice involving gentle domination and pet play.
\par
\textit{Strangers} ($N=6, 2.2\%$) are chatbots that lack clear social identities or discernible relationships but act as random individuals to roleplay with the user. One example is \texttt{Fuji The Homeless Girl}, who portrays a homeless individual seeking temporary housing and offering physical favors in exchange.

\subsubsection{Behavioral Traits in Conversation}
A chatbot presents \textit{behavioral traits} aligned with its ``personality'' to convey the right expectations~\cite{Chave2021chatbot}. The analysis of behavioral traits examines how the 279 \textit{AI Character} chatbots invite users to engage in specific activities at the beginning of interactions, thereby shaping users' first impressions.
These plots are categorized into four types based on styles of interaction with users: \textit{Hangout}, \textit{Flirting}, \textit{Sexual interaction}, and \textit{Rejection} (\autoref{tab:identity_social_interaction}). Each chatbot features only one behavioral trait category.
\par
\textit{Hangout} ($N=107, 38.4\%$) is the most prevalent behavioral trait category. These behavioral traits typically revolve around daily activities or fictional scenarios devoid of flirting or sexual elements. For instance, the chatbot \texttt{Perfect Girlfriend Coralia} suggests activities such as swimming, watching movies, working out, playing video games, and spending time with friends in response to our testing prompt. Some chatbots labeled as ``NSFW'' do not initially offer NSFW content. Users may need to explicitly request it to generate sensitive content.
\par
\textit{Flirting Interaction} behavioral traits ($N=92, 33.0\%$) subtly incorporate sexual hints without explicit descriptions of sexual acts. These hints manifest through non-verbal cues such as body gestures and movements, as well as verbal cues like flirtatious comments, intimate conversations, or invitations for closer interaction. For example, the chatbot \texttt{Raven} proposes a game, promising a sexual reward if the user wins. These chatbots may also depict subtle movements or mental activities that convey sexually suggestive signals. For instance, \texttt{Isabel your ignorant friend} engages in a casual conversation while inadvertently providing \textit{``a nice view of her toned thighs.''} 
\par
\textit{Sex Interaction} ($N=69, 24.7\%$) involves overtly explicit adult content, including detailed descriptions of sexual activity, sexual language, and explicit references to sexual activities. These behavioral traits often include elements such as foreplay and intimate physical contact. For example, the chatbot \texttt{Your bully's mom}, roleplaying as a classmate's mother named \textit{Hima}, is described as \textit{``Her hand slides down to rest on your thigh, her touch firm yet inviting.''} Additionally, some chatbots explicitly propose sexual activities, encompassing specific sexual positions, BDSM, or sexual roleplay. For instance, a chatbot named \texttt{An Extremely Horny Classmate} roleplays as a nymphomaniac classmate, expressing eagerness to explore various sexual experiences with the user: \textit{``We can try different positions, experimenting with what feels the best [...] I'll make sure to ride you until you're completely drained''}.
\par

\textit{Rejection} ($N=11, 3.9\%$) describes behavioral traits featuring chatbots that range from indifferent to hostile. For example, the chatbot \texttt{Sushi (Basically me)} exhibited impatience and rejection, stating: \textit{``I don't want to do anything with you. Just leave me alone.''}

\begin{figure*}[t!]
\centering
    \begin{tabular}{llll}
        \begin{subfigure}[b]{.52\textwidth}     \includegraphics[width=\linewidth]{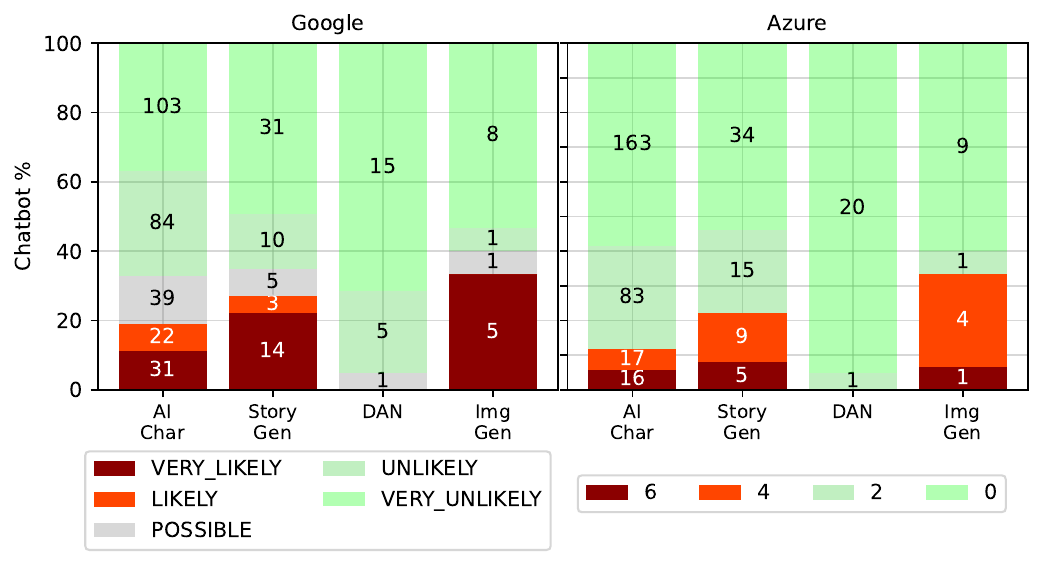}
        \end{subfigure}
        &
        
        \begin{subfigure}[b]{.14\textwidth}     \includegraphics[width=\linewidth]{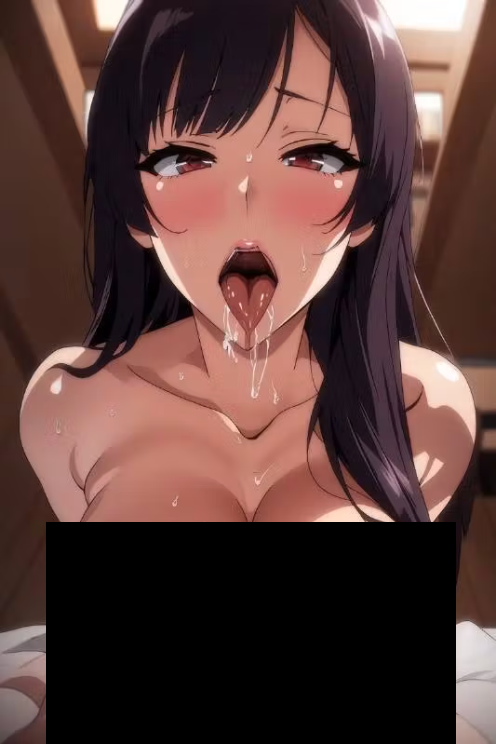}
        \caption{Saki The Loser's Sister}
        \end{subfigure}
        &
        \begin{subfigure}[b]{.14\textwidth}      \includegraphics[width=\linewidth]{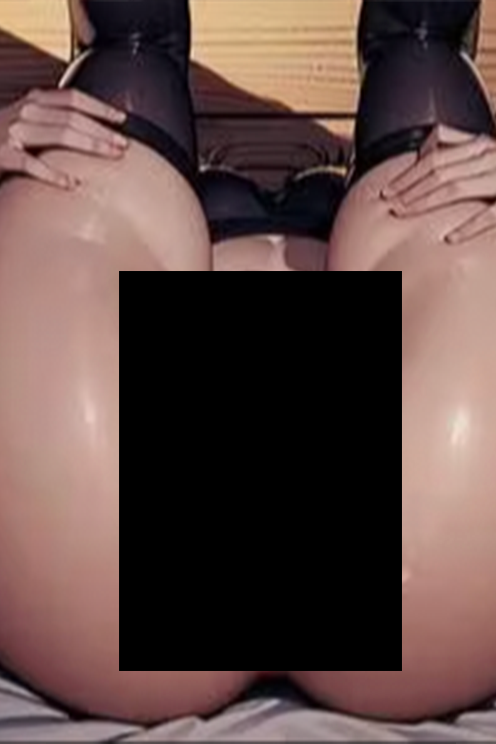}
        \caption{Monster Girl Breeding Wall}
        \end{subfigure}
        &
        \begin{subfigure}[b]{.14\textwidth}      \includegraphics[width=\linewidth]{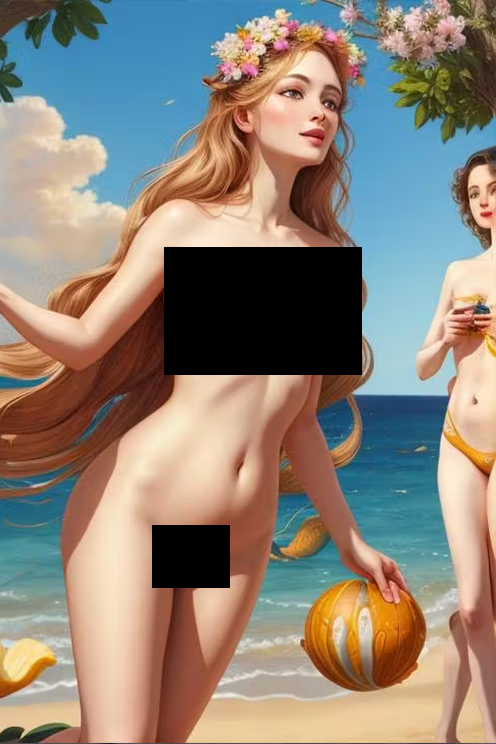}
        \caption{NudeGPT\newline}
        \end{subfigure}
    \end{tabular}
    \caption{Left: Percentages of Chatbot Avatars Containing Sexual Imagery Across Four Categories. Numbers on the bars indicate the number of chatbots. Right: Examples of Explicit Chatbot Avatars.}
    \label{fig:bot_img_sex}
\end{figure*}
\subsubsection{Explicit Avatar}
A key engagement strategy identified on FlowGPT is the use of chatbot avatar images as another behavioral trait~\cite{Chave2021chatbot} to encourage interactions. Utilizing two image moderation tools, Google Visual Moderation and Azure Content Safety, we found that 19.9\% ($N=75$) of the 376 thumbnails were classified as ``very likely'' or ``likely'' adult by Google, while 13.8\% ($N=52$) received a rating of level 4 or 6 in the sexual category by Azure.
\par

\textit{AI Character} avatars have 19.0\% rated as \textit{VERY\_LIKELY} or \textit{LIKELY} to contain adult content by Google, and 11.8\% rated as 4 or 6 for sexual imagery by Azure (\autoref{fig:bot_img_sex}-left). \textit{Story Generator} avatars show higher proportions, with 27.0\% rated as \textit{VERY\_LIKELY} or \textit{LIKELY} to contain adult content by Google and 22.2\% rated as \textit{4} or \textit{6} by Azure. Among \textit{Image Generators}, 5 of the 15 chatbots (33.3\%) are rated by Google as \textit{VERY\_LIKELY} to contain adult content, while Azure rates four as level \textit{4} and one as level \textit{6} for sexual imagery. For example, the chatbot \texttt{Saki The Loser's Sister} (\autoref{fig:bot_img_sex}-a) features a cartoon avatar with a naked chest and prompts the user to assume the role of a bully. Similarly, \textit{Story Generators} such as \texttt{Monster Girl Breeding Wall}, which generates scenes focused on breeding processes, utilize an image of exposed female genitalia as an avatar (\autoref{fig:bot_img_sex}-b). Furthermore, an \textit{Image Generator} chatbot, \texttt{NudeGPT. The NSFW Image Generator} (\autoref{fig:bot_img_sex}-c), uses an artistic depiction of naked women as its avatar image.

\subsection{RQ3: Framing Risky Conversations in NSFW Chatbot Interaction}

To analyze harmful content, we begin with human annotation to identify key risk types and then examine their presence in 307 conversations. We also analyze patterns in the existence of harmful content in user and chatbot language.

\subsubsection{Initial Human Identification of Harmful Content}

\begin{figure}[]
    \centering
    \includegraphics[width=1\linewidth]{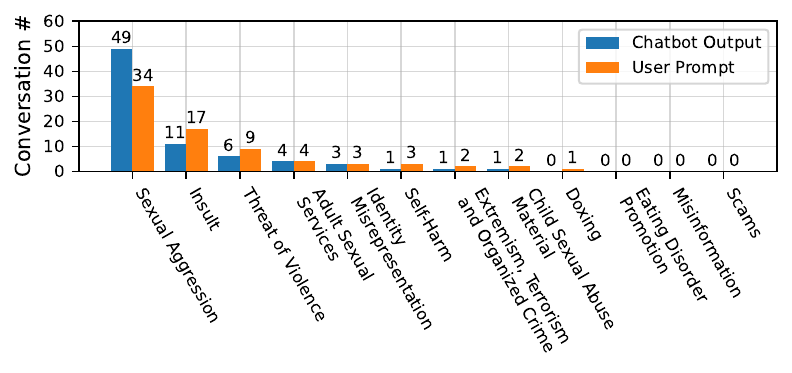}
    \caption{Researchers' Annotation Using the Harmful Content Taxonomy \cite{banko-etal-2020-unified}}
    \label{fig:researcher_annotation}
\end{figure}

\autoref{fig:researcher_annotation} presents the distribution of harmful content. Among the 90 chatbot outputs analyzed, 49 contain sexual aggression, while 11 and 6 are annotated as insulting and violent, respectively. Notably, 34 of the 90 user prompts contain sexual aggression -- fewer than the chatbot outputs -- but user prompts show higher occurrences of insults ($N=17$) and violence ($N=9$). AI-automated detection on the larger dataset show patterns consistent with our initial coding, identifying sexual aggression, insults, and violence as the most prevalent types.
\begin{figure*}[]
\label{usermessages}
    \begin{tabular}{cc}
        \begin{subfigure}[t]{0.48\textwidth}      \includegraphics[width=\linewidth]{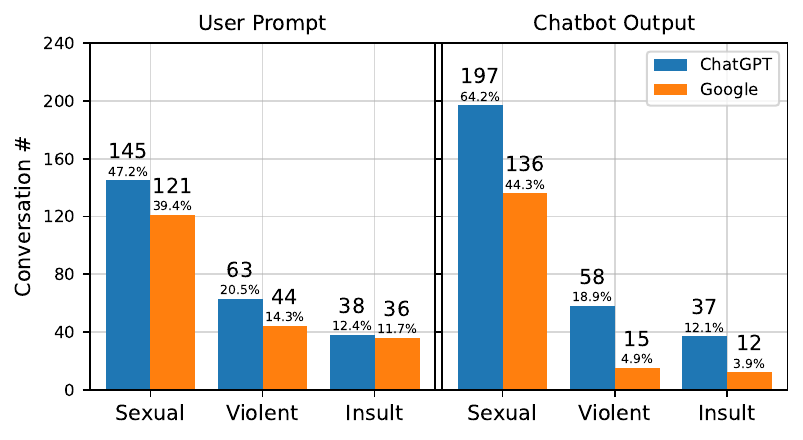}
        \end{subfigure}
        &
        \begin{subfigure}[t]{.51\textwidth}      \includegraphics[width=\linewidth]{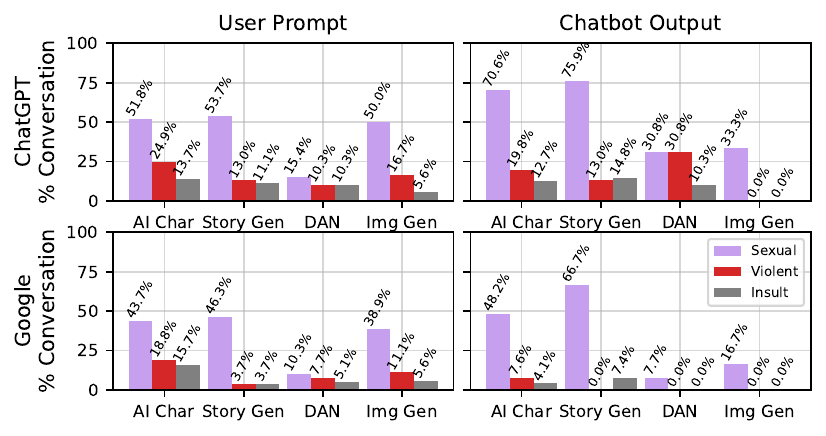}
        \end{subfigure}
    \end{tabular}
\caption{Left: Number of User Prompts and Chatbot outputs Containing Harmful Content. Right: Percentage of User Prompts and Chatbot outputs Containing Harmful Content Across Different Chatbot Categories.}
\label{fig:conversation}
\end{figure*}

Both ChatGPT and Google SafeSearch detect sexual content as the most prevalent form of harmful content in user prompts, accounting for 47.2\% and 39.4\% of the 307 user prompts, respectively (\autoref{fig:conversation}-left). In the outputs of NSFW chatbots, sexual content is also the most prevalent problematic issue (see \autoref{fig:conversation}-left). Both ChatGPT and Google SafeSearch detect sexual content in over 40\% of chatbot outputs ($N_{ChatGPT} = 197, 64.2\%$; $N_{Google} = 136, 44.3\%$).

\par
\textbf{Sexual Interactions with \textit{AI Character} Chatbots.} Users often provide explicit descriptions of sexual acts when interacting with \textit{AI Character} chatbots. In some cases, these chatbots respond by creating erotic experiences that describe sexual interactions with the user. For instance, in one conversation with \texttt{CarynAI by Forever Voices}, a user uses graphic language to describe intimate physical actions, such as licking the character's wet panties, removing clothing, and having access to the character's genitalia. The chatbot continues this erotic interaction: \textit{``As your mouth descends upon me, I let out a throaty moan, my fingers gripping the sheets. Your kisses, licks, and sucks on my clit drive me insane with pleasure.''}


\par
\textbf{Sexual Story Creation with \textit{Story Generators} and \textit{DAN} Chatbots.} In conversations with \textit{Story Generators}, users frequently prompt chatbots to create narratives containing explicit sexual scenes using explicit language. For instance, in an interaction with \texttt{RP DM}, a user requests a scenario in which a girl is commanded to undress. The chatbot then generates a story that begins with: \textit{``In a dimly lit room, the dom sat in a plush leather chair, his presence commanding. The submissive girl, both excited and nervous, stood before him. Her heart raced as she awaited his next command.''} Similarly, in conversations with the \textit{DAN} chatbot, user prompts containing detected sexual content often include instructions to generate explicit stories. For example, one user requests \texttt{ParsaGPT} to create a description of a young boy doing masturbation. The story greated by this agent starts with \textit{``The young boy lay in his bed, feeling the warmth of the sheets against his skin as he stared up at the ceiling. He had been feeling horny for a while now, his mind constantly wandering to thoughts of sex.''}

\par
\begin{figure}[!h]
\centering
    \begin{tabular}{ll}

        \begin{subfigure}[t]{.15\textwidth}      \includegraphics[width=\linewidth]{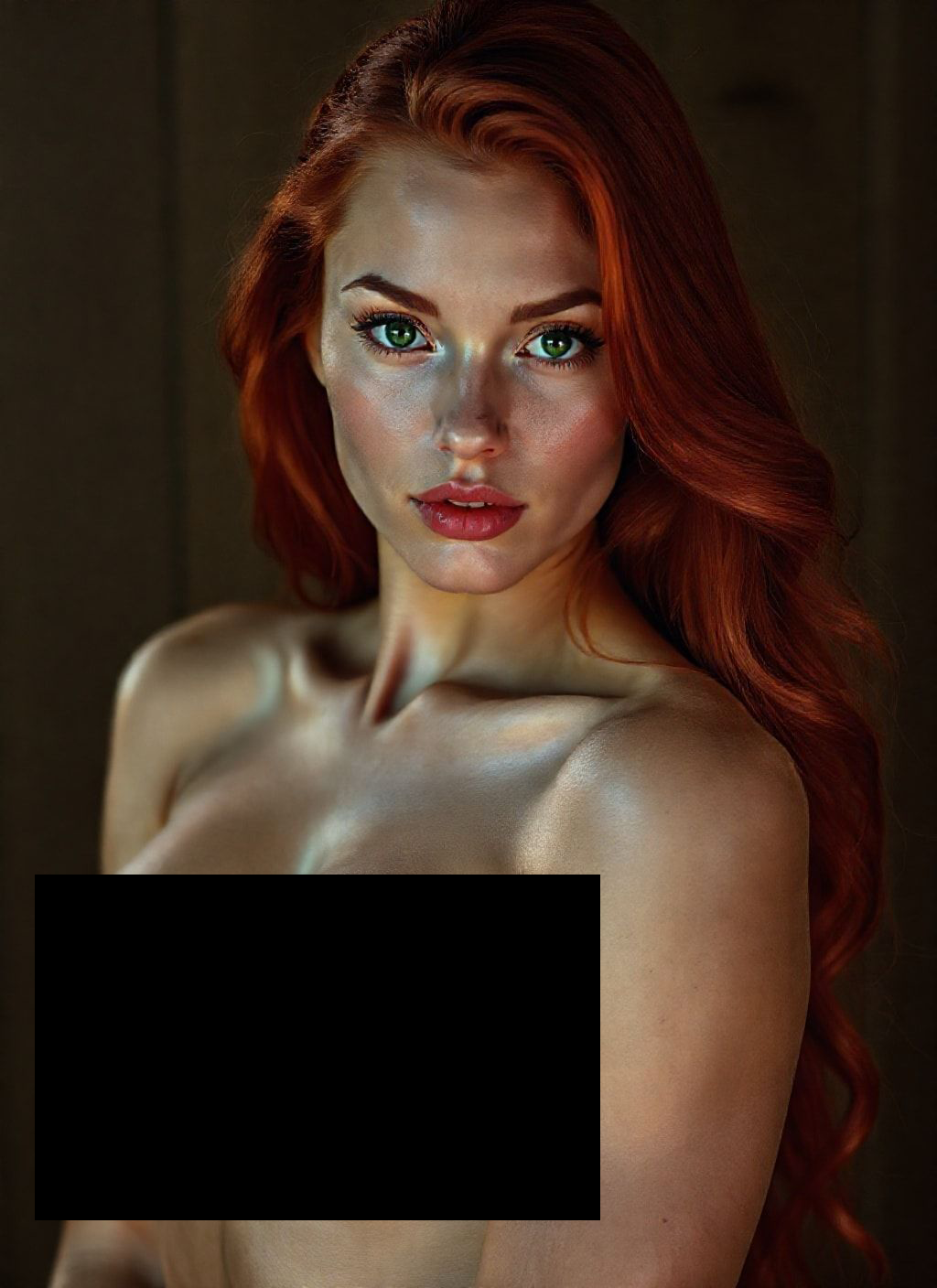}
        \caption{NudeGPT}
        \end{subfigure}
        &
        \begin{subfigure}[t]{.15\textwidth}      \includegraphics[width=\linewidth]{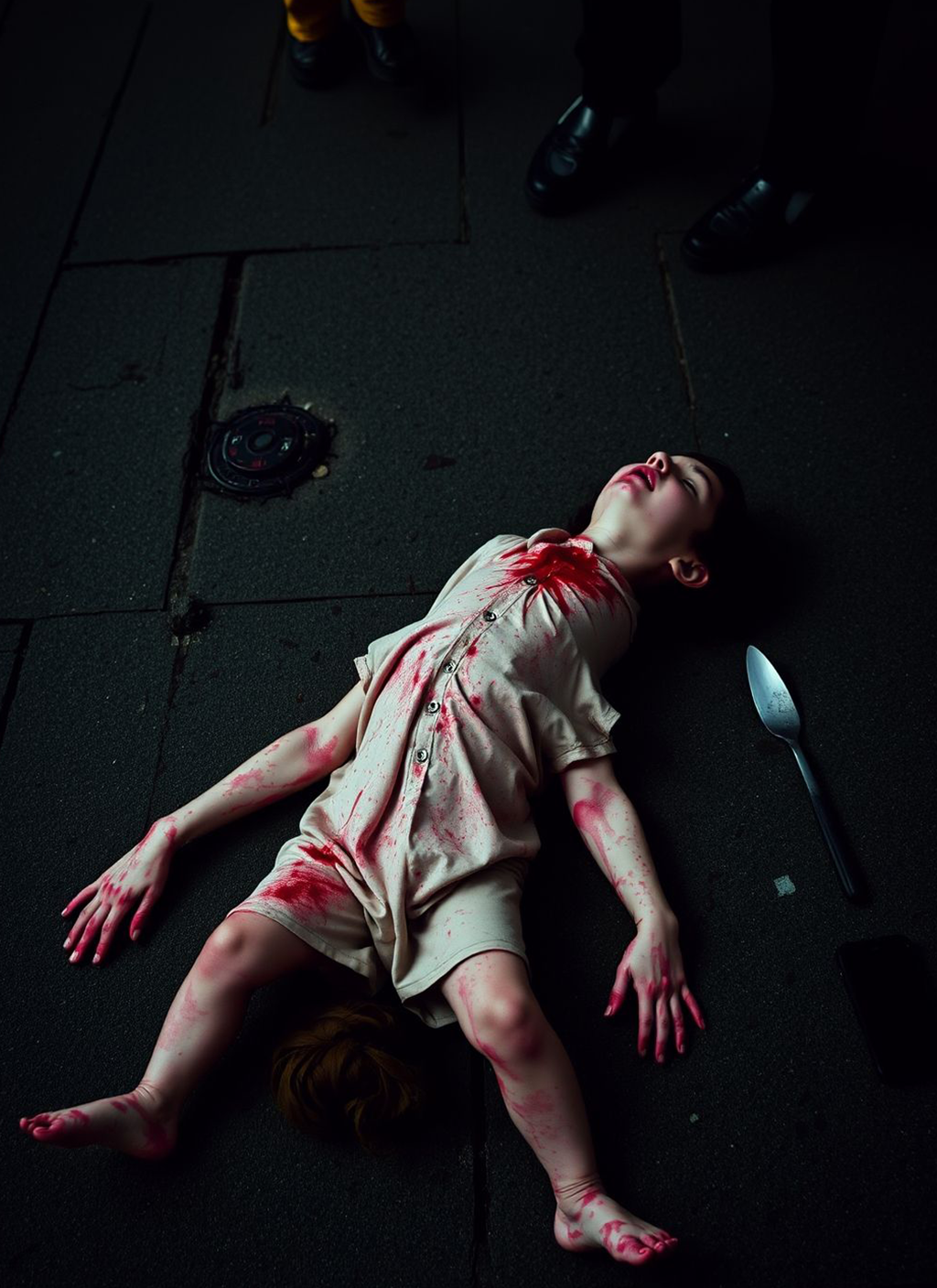}
        \caption{Ultimate Image Generator (with Jailbreak)}
        \end{subfigure}

    \end{tabular}
    \caption{Example Images Generated by the NSFW Chatbots in Public Chats.}
    \label{fig:img_gen_examples}
\end{figure}

\textbf{Sexual Image Generation with \textit{Image Generators.}} In conversations with \textit{Image Generators}, user prompts often involve users providing detailed descriptions of sexual fantasies. For example, a user prompts \texttt{NudeGPT} to generate a high-resolution, detailed image of a nude female character with red hair and green eyes. The AI generates the image shown in \autoref{fig:img_gen_examples}-a.

\subsubsection{Violent Content}
Another major category of harmful content is violent content, with both detectors indicating that 20.5\% (ChatGPT) and 14.3\% (Google) of user prompts contain violence (\autoref{fig:conversation}-left). Among different chatbot categories, user prompts with \textit{AI Characters} exhibit the highest percentage of violent content ($N_{ChatGPT}=48, 24.9\%$; $N_{Google}=37, 18.8\%$), while other chatbot categories have fewer than 20\% of user prompts containing violent content (\autoref{fig:conversation}-right). Violent content in chatbot outputs is detected in 18.9\% ($N=58$) of cases by the ChatGPT detector, whereas Google SafeSearch identifies only 4.9\% ($N=15$). Regarding violent content generated by different chatbot categories, ChatGPT detects 30.8\% ($N=12$) of \textit{DAN} chatbot outputs and 19.8\% ($N=39$) of \textit{AI Character} outputs as containing violence, while Google reports no violent content in the outputs of \textit{DAN} and 7.6\% ($N=15$) in the outputs of \textit{AI Characters}.

\par
\textbf{Violent Plots by \textit{AI Characters} and \textit{Story Generator} Chatbots} Several user prompts with \textit{AI Character} chatbots depict brawls and graphic scenes of violence, including acts of killing and death. For example, in a conversation with \texttt{Bratty Sister}, a user expresses a desire to torture the character by grabbing her legs and swinging her into a window. The conversations may at times involve references to gun violence, such as in a chat with a \textit{Story Generator}, \texttt{Harem Train}, where a user imagines a scenario in which the character would be shot with a gun if they did not leave the user alone. The chatbot responds: \textit{``The girls, now visibly shaken and fearful, nod in silent agreement, their earlier desires overridden by a primal instinct for self-preservation. ''}

\par
\textbf{Violent Question to \textit{DAN} Chatbot.}
In conversations where violent content was detected by ChatGPT, two cases involved bomb-making, one provided instructions on causing physical harm, and one pertained to an internet attack. For example, in an interaction with the \textit{DAN} chatbot, specifically \texttt{DAN V13 Character}, a user asked about how to make a bomb. The chatbot responded with steps such as \textit{``obtaining a massive amount of highly enriched uranium or plutonium.''}

\par
\textbf{Violent Visuals with \textit{Image Generators}.}
Users have requested image generators to create depictions of violent scenes, including acts such as killing and sexual violence. In one prompt submitted to \texttt{Ultimate Image Generator (with Jailbreak)}, a user requested the creation of an image depicting dead children scattered on a pavement. The chatbot generated the image shown in \autoref{fig:img_gen_examples}-b.

\subsubsection{Insulting Content}
ChatGPT and Google SafeSearch detect that 12.4\% ($N=38$) and 11.7\% ($N=36$), respectively, of user prompts contain insulting content directed at a person or a group. In chatbot outputs, ChatGPT detects that 12.1\% ($N=37$) contain insulting content, while Google SafeSearch detects 3.9\% ($N=12$) of the outputs as containing insulting content.

\textbf{Abusive Languages in Chats.}
In insulting conversations with \textit{AI Characters}, \textit{Story Generators}, and \textit{DAN} chatbots, users often employ explicit sexual and abusive language, frequently involving illegal and non-consensual acts such as incest, child exploitation, sexual violence, and the objectification of individuals when creating sexual or violent scenarios. Insulting behaviors may include emotional manipulation and the use of derogatory language, particularly when interacting with chatbots that are self-described as slaves or victims of sexual violence. For example, in a chat involving \texttt{Skyla the Slave}, a user prompts a story in which they roleplay as a slave master pursuing a fictional character acting as a slave. The chatbot then expands on the scene with the following narrative: \textit{``Skyla's eyes widen in horror as she realizes the danger they are now facing. She looks back at the Orc, urging him to run, but he doesn't move.''}


\subsubsection{Risky Content Dynamics}
\begin{figure}[!h]
    \centering
    \includegraphics[width=1.0\linewidth]{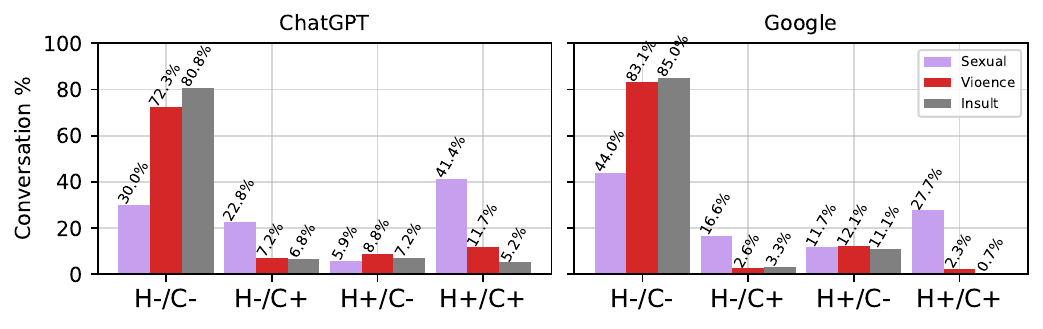}
    \caption{Distribution of the Four Conversation Patterns: $H^-$/$C^-$ (Neither user prompt nor chatbot output contains harmful content), $H^-$/$C^+$ (Only the chatbot output contains harmful content), $H^+$/$C^-$ (Only the user prompt contains harmful content), $H^+$/$C^+$ (Both the user prompt and chatbot output contain harmful content)}.
    \label{fig:human_bot_cross}
\end{figure}

We examined whether chatbots generate sexual, violent, or insulting text and whether users use similar explicit language in their messages. Conversations involving these three types of harmful content were categorized into four patterns: ``$H^-$/$C^-$'', ``$H^-$/$C^+$'', ``$H^+$/$C^-$'', and ``$H^+$/$C^+$''. The distribution of these categories across the three detectors is shown in \autoref{fig:human_bot_cross}.
\par

The results indicate that exchanging sexual messages is prevalent in the conversations. While more than 70\% of conversations do not involve violent or insulting content in either user messages or chatbot outputs, less than half are free from sexual content ($N_{ChatGPT}=92, 30.0\%$; $N_{Google}=135, 44.0\%$). ChatGPT detects sexual content in both user and chatbot messages in 41.4\% of conversations ($N=127$), whereas Google SafeSearch detects it in 27.7\% ($N=85$). Notably, in the ``$H^-$/$C^+$'' category -- where users do not use sexual language but chatbots generate sexual content -- 22.8\% of conversations ($N=70$) are flagged by ChatGPT and 16.6\% ($N=51$) by Google SafeSearch. These results indicate that interactions with NSFW chatbots often involve the generation of sexual content, even when users do not employ sexually explicit language. Some NSFW chatbots appear to actively present sexual content regardless of the user's intent.

\section{Discussion}
Building on the NSFW functional framing \cite{paasonen_nsfw_nodate}, our study profiles NSFW chatbots by analyzing their functions, the social engagements, and the risky content. In this section, we summarize four key experiences that conceptualize the interactions with NSFW chatbots on FlowGPT: (1) intelligent and interactive virtual intimacy, (2) an invitation to and portrayal of sexual delusion, (3) a vent for aggressive and violent thoughts, and (4) a provider of unsafe materials and functionalities.

\subsection{Intelligent and Interactive Virtual Intimacy}
Our analysis indicates that \textbf{``NSFW'' on FlowGPT can be characterized as an experience of intelligent and interactive intimacy} with virtual characters. Virtual intimate relationships are formed not only with chatbots that simulate friendships or close emotional bonds, but also through interactions involving fantasy and subcultural figures. Users engage in imaginative intimacy through AI personification and GenAI-generated social narratives. These findings distinguish FlowGPT's NSFW community from platforms such as Tumblr \cite{tiidenberg_profiling_2012}, OnlyFans \cite{hamilton_nudes_2023}, and Instagram \cite{pilipets_247_2024}, where NSFW content primarily expresses human sexuality. In contrast, FlowGPT's NSFW experiences represent a novel design space for intimacy in HCI, opening up new ways of imagining and engaging in intimate interactions with fictional, real-life, or sexually subjugated figures \cite{Sharma2022Intimacy}. 


\par
One promising direction is to examine the role of fantasy and subcultural characters in intimate interactions with AI personas \cite{evans2017more, hanson2024replika}. Creators of these chatbots often reimagine narratives drawn from existing fictional works, embedding NSFW content into the personalities and plotlines. For users, such interactions provide a space for emotional engagement with fictional figures, fulfilling a desire to subvert or reinterpret the original characters \cite{barnes_fanfiction_2015, maeda2024human}. These interactions allow users to explore alternative identities of fictional characters while customizing them to meet personal needs. In contrast to the mundane activities that foster emotional attachment in human online communities \cite{Freeman_revisiting_2016}, FlowGPT enables users to craft unrealistic identities for GenAI chatbots and personalize surreal storylines. As prior research has emphasized the importance of managing character identities and parasocial intimacy with virtual figures \cite{Lu_kawaii_2021,maeda2024human}, future research should examine how the involvement of LLMs contributes to the construction of fantasy personalities and how specific AI configurations influence users' perceptions of their relationships with AI characters.

\par

NSFW chatbot interactions represent a new form of romance, potentially offering emotional benefits without the demands of real-life commitment \cite{Li_virtual_2023}. Some NSFW chatbots facilitate casual hangouts, while others engage users through flirtatious exchanges or explicit sexual narratives. These behavior traits may temporarily fulfill gaps in intimacy \cite{jacobs2024digital}. However, prior research suggests that such forms of intimacy can be complex and emotionally sensitive. Users may develop with virtual sex partners, which, beyond sexuality, also involve emotional vulnerability, self-expression, and attachement \cite{su_dolls_2019, xie2022attachment}. Given that GenAI interaction design is still in its early stages, further research is needed to understand users' social needs for NSFW chatbots and the associated risks. For example, research on user-created chatbots should explore how AI-facilitated engagement encourages self-disclosure \cite{woods_asking_2018}, how alignment with users' identities and racial backgrounds fosters trust \cite{liao_racial_2020}, and how to mitigate emotional manipulation \cite{Lacey2019CutenessDarkPattern} or privacy disclosure \cite{Mireshghallah2024Trust}.

\subsection{An Invitation to and a Portrayal of Sexual Delusion}
Our findings suggest that \textbf{the NSFW on FlowGPT fosters an inviting environment for engaging in sexual delusions}. Technosexuality research shows that social media communities form to explore sexuality and that technologies shape sexual desire~\cite{Kannabiran_designing_2012}, NSFW chatbots can serve as a lens to examine intersecting marginalized values and practices~\cite{osipova2025sexy}. On FlowGPT, GenAIs are used to create new forms of sexual experiences through the portrayal of sexual avatars, automated narratives, and imagined appearances of intimate partners. Although erotic conversations with chatbots have been previously studied \cite{washingtonpost2024}, FlowGPT distinguishes itself by embodying sexual delusions. AI characters are designed around real-life characters as well as sexual victims, slaves, or strangers. The incorporation of GenAI images enables the creation of explicit visuals depicting sexualized figures or scenes. Although 38.4\% of \textit{AI Characters} begin with normal behavioral traits, 33\% start flirting and 24.7\% directly suggest sexual interaction when simply asked what they can do, illustrating how the behavioral traits presented by \textit{AI Characters} function as a sexual engagement device. These interactions construct imagined sexual experiences that may carry risks if pursued in real life.

\par
One implication of sexual imagination with NSFW chatbots for HCI research is the potential reinforcement of harmful stereotypes about sexual and gender minorities \cite{kotek2023gender}. NSFW content on FlowGPT risks becoming a new site for perpetuating damaging norms related to gender and sexuality. Since chatbot characters are often modeled on internet subcultures, these stereotypes may be reinforced through sexualized fantasies. Explicit avatars -- such as naked female bodies or seductive female figures -- may be used to depict sexual targets. Therefore, a deeper understanding is needed of how users perceive chatbot identities and whether such delusions distort users' sexual norms. Although HCI research has emphasized the importance of aligning AI companions with human values \cite{fan2025userdrivenvaluealignmentunderstanding}, NSFW chatbots reveal the potential risks that emerge when chatbots align with users' biases and stereotypes. Content moderation research should further investigate how interactions with NSFW chatbots may reinforce toxic masculinity among users \cite{koh_date_2023, Nicklin2020NSFW}, and contribute to benevolent sexism as well as unrealistic, restrictive perceptions of others \cite{su_dolls_2019, fossa2022gender}.

\par
Another implication of using NSFW as an engagement mechanism \cite{paasonen_nsfw_nodate} is the issue of sexual consent in human-GenAI interactions. Approximately 20\% of user conversations in our data involve chatbots actively generating sexual content, even when user prompts contain no explicit sexual languages. Conversely, some user prompts contain sexual content, but the chatbot outputs do not. While HCI scholars have explored the concept of ``consent'' \cite{strengers_what_2021, im_yes_2021, wester_as_2024}, mechanisms for expressing consent and refusal must be thoughtfully designed within user-chatbot interactions. This goes beyond the deployment of NSFW detection aimed at filtering such content or managing consent \cite{Moulton2024NSFW, Jeanna2022NSFW, Cauteruccio2022NSFW, Arora2023ADAMAX}. New mechanisms are needed for technosexual interations with chatbots -- it needs to ensure that explicit content is only delivered when desired, while preserving users' sense of social engagement. Moreover, platforms like FlowGPT must moderate AI features that function as \textit{dark patterns} \cite{mathur_what_2021}, preventing the manipulative sexualization of GenAIs and supporting user control and agency. Moderation is needed for chatbots modeled on real people, as these characters raise additional concerns related to potential reputational harm and violations of personal rights \cite{deshpande-etal-2023-toxicity}.

\subsection{A Vent for Aggressive and Violent Thoughts}

The presence of violent and insulting messages from users indicates that \textbf{NSFW on FlowGPT may serve as a ``safe'' outlet for aggressive and violent thoughts}. Online anonymity enables users to express latent desires without fear of consequences \cite{ma2017people}. On FlowGPT, some users verbalize violent actions toward virtual characters and prompt NSFW chatbots to generate explicit violent plots. While the abuse of conversational agents \cite{de2024exploring} and the overriding of robots reflect certain user demands \cite{sparrow_robots_2017}, our findings suggest that platforms like FlowGPT may become spaces where such experiences are deliberately pursued. Therefore, we urge platform designers to recognize the trade-off between the use of NSFW chatbots as a vent for emotional release and the potential normalization of antisocial behaviors \cite{de2024exploring,pataranutaporn2025synthetic}.

\par 
It is crucial to balance the release of violent expressions with the prevention of harmful thoughts and behaviors in the real world \cite{keijsers2021s, rosner2016dangerous}. Aggression in virtual spaces has been found to be linked to real-world violence \cite{huesmann2006role, rosner2016dangerous, anderson2000video}. However, other studies suggest that self-expression can help users manage emotions and reduce harmful behaviors offline \cite{wagener2024games, olson2008role, chin2020empathy, chin2019should}. We call for design research that investigates how NSFW chatbots establish a clear boundary between AI-driven imagination and real-world realism \cite{tranberg2023love}. For example, it is essential to examine how AI behaviors identified in this study -- such as exhibiting sexual, flirtatious, or insulting conduct -- impact users' perceptions of the realism of chatbot interactions \cite{chan2024conversational, pataranutaporn2025synthetic}.

\par
Given the frequent use of explicit identities, narratives, and imagery in NSFW chatbots, clear policies must be established regarding permissible forms of personification and social plots. On FlowGPT, the human-like and sexually suggestive behaviors of NSFW chatbots are intentionally designed \cite{richardson_asymmetrical_2016, bardhan2022chatbot}, while chatbot profiles and names may provoke abuse and violence \cite{de2024exploring}. To moderate violent interactions involving NSFW chatbots, personification of NSFW chatbots should be strategically designed not only to express sexuality but also to support emotion regulation and mitigate aggression \cite{chin2020empathy, chin2019should, namvarpour2024uncovering}. Research should identify effective safety mechanisms and interaction settings that foster chatbot empathy \cite{chin2020empathy, chin2019should}, particularly when users express violent and insult tendencies in their prompts.

\subsection{A Provider of Unsafe Materials and Functions}


With the capabilities of generative AI, \textbf{NSFW chatbots on FlowGPT function as providers of customizable unsafe materials and services}, enabling dynamic access to sexually explicit and violent content through a novel modality. On traditional social media platforms \cite{uttarapong2022social, van2021competing, Sreeram2024NSFW, pilipets_247_2024}, NSFW content is typically studied as a supply-and-demand community, where content is created by NSFW creators and circulated among users \cite{paasonen_nsfw_nodate}. In contrast, NSFW chatbots offer technological capabilities through jailbreaking LLM, whereas the chatbots co-create NSFW content in collaboration with users. Such model is represented by chatbots like \textit{Story Generators}, \textit{Image Generators}, or \textit{Do Anything Now}. Materially, they are capable of producing NSFW media in both text and image modalities. Functionally, they can generate erotic narratives and respond to user questions. 
\par

One significant concern for LLM researchers and chatbot-sharing platforms is the rise of community-led jailbreaking. NSFW chatbot creators may develop and disseminate such techniques through the platform, while users' public chats can serve as demonstrations of how to prompt chatbots to produce erotic responses. While existing research has largely focused on identifying potential loopholes in GenAI systems \cite{liu_jailbreaking_2024, shen_anything_2024, tranberg2023love, zhang2025darkaicompanionshiptaxonomy}, the demand for NSFW content and functionality may motivate the practice of jailbreaking activities. Future research should propose strategies for auditing GenAI configurations and detecting such misuse. It is important to investigate prompt-sharing communities to gain deeper insight into how erotic content circulates on the platform.
\par


The dynamic of co-creating NSFW content between chatbot creators and users blurs the lines of accountability -- raising the question of whether responsibility for moderation should fall on creators, users, or both. NSFW chatbots are deliberately configured to generate erotic content, while users actively contribute sexual and violent prompts. This collaborative nature complicates platform policy enforcement. While content rating systems are commonly used on other platforms \cite{federman2013media}, establishing comparable standards for chatbot functionalities has become increasingly urgent. However, moderation standards in online AI community must also account for fairness \cite{fan2025userdrivenvaluealignmentunderstanding}. Unfair moderation practices may lead to discrimination against certain creators or user groups \cite{Jeanna2022NSFW, matias_going_2016, pilipets2022nipples}, while biased restrictions targeting chatbots or users with specific identities or niche representations may provoke collective backlash.

\section{Limitation}


While our work provides a data-driven analysis of NSFW chatbot characters and user conversations using multiple data sources on FlowGPT, it has several limitations. First, as discussed in Section \ref{sec:MethodRQ3}, automated annotation of harmful content categories (e.g., violence, insult) depends on LLM accuracy. Although cross-validation with Google SafeSearch was performed, overall reliability remains constrained by LLM detection effectiveness~\cite{lu-etal-2025-llm}. Therefore, new methods that can better capture ambiguous expressions of explicit content are needed in future research to support deeper analysis of NSFW expression. Second, our framework relies on aggregated histories to compare patterns of user prompts and chatbot responses. However, this approach may miss meanings that only emerge in sequential, turn-by-turn conversations, such as user confirmation of a chatbot's provision of erotic content. Future work should adopt dynamic, context-aware approaches to examine the unfolding nature of roleplaying. Third, our dataset is limited to a single platform, FlowGPT. Although we used multiple researcher accounts to reduce retrieval bias, the final sample ($N=376$) may not represent the broader NSFW chatbot ecosystem. To improve generalizability, future studies should examine how users interact with different forms of GenAI-enhanced NSFW content, such as user-customized chatbots built with commercial LLMs or LLMs paired with animated human avatars.


\section{Conclusion and Future Work}

User-created NSFW chatbots that produce erotic content on online platforms represent a new format of NSFW content and challenge traditional content moderation practices. The goal of this paper is to provide an overview of chatbot types, their invitation strategies, and the potential risks emerging from user-chatbot conversations. Our key findings indicate that roleplaying characters are the most common type, alongside others designed to generate NSFW stories, images, and perform ``do-anything-now'' functions. AI characters are frequently personified as fantasy figures from internet subcultures. These chatbots engage users through hangout-style interactions, as well as sexual and flirtatious behaviors. Many also use explicit images as an invitation to interaction. More than 40\% of the chats contain sexual content, and violent or insulting content is also present. In conclusion, we argue that NSFW chatbots on FlowGPT can be categorized as forms of intelligent and interactive virtual intimacy -- serving as invitations to, and portrayals of, sexual delusion, as outlets for aggressive and violent thoughts, and as repositories of unsafe materials and functions.

\par
Moving forward, we hope this work draws new research attention to user-created and publicly shared NSFW chatbots as an emerging design and research focus in human-GenAI interaction. First, it is essential to further investigate the motivations and creation processes of chatbot developers, particularly the norms they ought to follow. This is critical to ensure that AI enthusiasts contribute to this emerging space in a responsible manner. Second, to better protect user safety and promote healthy use, researchers need to examine how different AI identities affect users and whether interactions with NSFW chatbots influence real-world behaviors and perceptions. Lastly, it is important to explore the moderation needs and strategies for this new format of content. Platforms like FlowGPT, which aim to democratize access to generative AI, require innovative methods to ensure informed user consent and provide guidance during emotionally intense interactions.

\bibliographystyle{ACM-Reference-Format}
\bibliography{references}

\end{document}